\documentclass[reprint,amsmath,amssymb,aps,nofootinbib]{revtex4-1}

\usepackage{amssymb,amsmath,amsfonts}
\usepackage{graphicx}
\usepackage{dcolumn}
\usepackage{bm}
\usepackage{hyperref}
\usepackage{color}
\usepackage{url}
\usepackage{xfrac}
\usepackage[letterpaper,margin=1.0in]{geometry}
\usepackage{epsfig}
\usepackage[normalem]{ulem} % this package is just for adding the cross line
\usepackage{caption}
\usepackage{subcaption}
\usepackage{multirow}
\newcommand{\diff}{{d}}
\DeclareMathOperator*{\argmin}{arg\,min}

\newcommand{\calE}{{\mathcal{E}}}
\newcommand{\calL}{{\mathcal{L}}}

\begin{document}

\title{The Affine Wealth Model:\\
An agent-based model of asset exchange that allows for \\
negative-wealth agents and its empirical validation %\\
% \textcolor{red}{\normalsize\bf Draft Copy, Do Not Circulate}
}
\thanks{\copyright 2018, all rights reserved}

\author{Jie Li}
\affiliation{Department of Mathematics, Tufts University, Medford, Massachusetts 02155, USA}
\author{Bruce M.\ Boghosian}
\affiliation{Department of Mathematics, Tufts University, Medford, Massachusetts 02155, USA}
\author{Chengli Li}
\affiliation{Department of Mathematics, Tufts University, Medford, Massachusetts 02155, USA}

% \date{\today}

\begin{abstract}
We present a stochastic, agent-based, binary-transaction Asset-Exchange Model (AEM) for wealth distribution that allows for agents with negative wealth.  This model retains certain features of prior AEMs such as redistribution and wealth-attained advantage, but it also allows for shifts as well as scalings of the agent density function.  We derive the Fokker-Planck equation describing its time evolution and we describe its numerical solution, including a methodology for solving the {\it inverse problem} of finding the model parameters that best match empirical data.  Using this methodology, we compare the steady-state solutions of the Fokker-Planck equation with data from the United States Survey of Consumer Finances over a time period of 27 years.  In doing so, we demonstrate agreement with empirical data of an average error less than 0.16\% over this time period.  We present the model parameters for the US wealth distribution data as a function of time under the assumption that the distribution responds to their variation adiabatically.  We argue that the time series of model parameters thus obtained provides a valuable new diagnostic tool for analyzing wealth inequality.
\end{abstract}

\maketitle

%\tableofcontents

\section{Introduction}
\label{sec:Introduction}

Stochastic agent-based Asset-Exchange Models (AEMs) for wealth distribution were first introduced in 1986~\cite{bib:Angle}, and first analyzed mathematically using methods of kinetic theory in 1997~\cite{bib:IspolatovKrapivskyRedner}.  AEMs are highly idealized but nonetheless very useful models for understanding the statics and dynamics of wealth distributions, and they exhibit a rich phenomenology~\cite{bib:Yakovenko, bib:ChakrabortiChakrabarti}.  In this work, we shall demonstrate that, with appropriate features included, they are also capable of explaining certain empirical economic data with remarkable accuracy.

Though there are many variants of AEMs, most consider a closed economy involving a fixed number of economic agents, each possessing a certain amount of a resource, which we shall refer to as {\it wealth}.  Pairs of agents are  chosen randomly to engage in binary transactions in which a small amount of wealth, $\Delta w$, might move from one agent to the other.  In the limits of large populations and long times, the agent density function may be described by a continuous distribution, either classical or singular in nature.  While such models can be easily simulated using the Monte Carlo method, in the limit of large numbers of small transactions -- i.e., in the so-called {\it small-transaction limit} -- partial differential equations of Fokker-Planck type~\cite{bib:Fokker, bib:Planck, bib:Kolmogorov} can be derived for their agent density functions.  These facilitate the study of the properties of these distributions, both analytically and by numerical methods.
 
The present paper is based on a particular AEM which was first proposed in 2002~\cite{bib:Chakraborti2002}, and got its name the {\it Yard-Sale Model} (YSM) later in the same year~\cite{bib:Hayes}. This model posits binary transactions in which $\Delta w$ is proportional to the wealth of the poorer of the two agents, and in which the direction that the increment of wealth moves is determined by a fair coin flip.  Numerical simulations at the time indicated that its time-asymptotic state was one of complete wealth condensation~\cite{bib:BouchaudMezard2000,bib:Burda2002}, or {\it oligarchy}, in which all the wealth falls into the hands of a single agent.  In 2014, a Boltzmann equation was derived for this model, and shown to reduce to a Fokker-Planck equation in the small-transaction limit~\cite{bib:Boghosian2014a,bib:Boghosian2014b}.  A year later it was definitively proven that the time-asymptotic limit of the evolution of either of these equations, from any initial condition, was the state of complete oligarchy~\cite{bib:Boghosian2015}.

In an effort to avert the state of complete oligarchy and thereby make the YSM more realistic, a model of redistribution was introduced whereby a flat wealth tax $\chi$ per unit time was imposed on all agents on a per-transaction basis, and redistributed uniformly to the entire population.  It was shown that this has the effect of adding an Ornstein-Uhlenbeck~\cite{bib:OU} term to the Fokker-Planck equation~\cite{bib:Boghosian2014b}.  A subsequent extension introduced the idea of Wealth-Attained Advantage (WAA) to the model, in order to account for the well documented privileges that wealthier agents enjoy over poorer agents, such as higher returns on investment and lower interest rates on loans~\cite{bib:Boghosian2017}.  Mathematically, this was accomplished by biasing the coin in favor of the wealthier agent.  The amount of the bias was taken to be proportional to the difference in wealth between the richer and poorer agent, times a coefficient, $\zeta$.  Because the bias is proportional to the wealth difference, it naturally reduces to zero when the transacting agents have equal wealth.

The resulting Extended Yard-Sale Model (EYSM)~\cite{bib:Boghosian2017}, with redistribution rate $\chi$ and WAA parameter $\zeta$, admits much more interesting phenomenology, which we review in Subsection~\ref{ssec:EYSM}.  The agent density function is a classical distribution in the {\it subcritical} regime defined by $\zeta < \chi$.  When $\zeta\geq\chi$, this passes to a partially wealth-condensed {\it supercritical} regime, characterized by coexistence between a classical distribution of agents with a fraction $\sfrac{\chi}{\zeta}$ of the total wealth, and an oligarchy with the remaining $1-\sfrac{\chi}{\zeta}$ of the total wealth~\cite{bib:Boghosian2017, bib:Boghosian2016}.  Further study revealed the presence of a ``duality'' between the subcritical and supercritical phases of this extended model~\cite{bib:LiBoghosian2017}.

In this paper, we introduce a further modification to the EYSM to allow for agents with negative wealth.  Such agents are widely observed in empirical data -- for example, approximately 10.9\% of the population of the United States in 2016 had liabilities in excess of assets, and hence negative wealth -- but most prior AEMs, including the EYSM, have restricted wealth to be a positive quantity.  Realistic models for agent density functions possess invariance properties under scalings of the total number of agents and the total wealth; we accomplish the extension to negative wealth by additionally requiring invariance of the wealth distribution under additive shifts.  Because the new model is invariant under both scalings and shifts, we refer to it as the {\it Affine Wealth Model} (AWM).

We derive the Fokker-Planck equation obeyed by the agent density function for the AWM, and we describe its numerical solution for steady-state wealth distributions in the subcritical regime (without an oligarchy).  We explain how numerical solutions in the supercritical regime (with an oligarchy) can be obtained from their subcritical counterparts by exploiting the above-mentioned duality~\cite{bib:LiBoghosian2017}.
  
Finally, we present a detailed comparison of the results of the AWM with empirical data.  In particular, we compare the steady-state solutions of the Fokker-Planck equation with data from the United States Survey of Consumer Finances~\cite{bib:SCF2013} over a time period of 27 years.  In doing so, we demonstrate both that (i) each of the extensions that we introduced in the basic AEM resulted in improved fitting, and (ii) of all these models the AWM is the one most faithful to the empirical data.  Additionally, we present fitting parameters for the U.S.\ wealth distribution data as a function of time, under the assumption that the wealth distribution responds to their changes adiabatically.  We argue that this time series of model parameters provides a new way to extract useful information about wealth inequality in an economy.  As an example, we demonstrate a precisely defined way of quantifying the extent to which the U.S.\ wealth distribution is partially wealth-condensed.

\section{Background}
\subsection{The Extended Yard-Sale Model}
\label{ssec:EYSM}

In this subsection, we provide a brief description of the Extended Yard-Sale Model (EYSM) on which the current work is based.  A more complete description can be found in a reference~\cite{bib:Boghosian2017}.  We suppose that our population has $N$ agents, each with some positive amount of wealth, and collectively possessing total wealth $W$, so that the average wealth of an agent is $W/N$.  We suppose that a specific Agent 1 with wealth $w$ is transacting with a randomly selected Agent 2 with wealth $x$ at time $t$.  In the course of this transaction, the wealth of Agent 1 is increased by $\Delta w$, while that of Agent 2 is decreased by $\Delta w$, so that total wealth is conserved.  We suppose that $\Delta w$, which may be either positive or negative, is described by the statistical process
\begin{equation}
\Delta w = \sqrt{\gamma \Delta t}\; \min (w,x) \eta + \chi \left( \frac{W}{N} - w \right) \Delta t.
\label{eq:rw1}
\end{equation}

The leading term in Eq.~(\ref{eq:rw1}) is a small proportion of the wealth of the poorer of the two agents, since it is plausible that most people do not stake a large fraction of their wealth on a single transaction.  Because of this requirement, an agent's wealth can never become nonpositive, and the support of the agent density function is $(0,\infty)$.  The coin flip is modeled by the random variable $\eta$, which is positive if the wealth is moving from Agent 2 to Agent 1, and negative if it is moving from Agent 1 to Agent 2.  We model WAA by supposing that the expectation value of $\eta$ is given by
\begin{equation}
E[\eta] = \zeta \sqrt{\frac{\Delta t}{\gamma}} \left(\frac{w-x}{W/N}\right),
\label{eq:rw2}
\end{equation}
where $\zeta$ is called the {\it WAA coefficient}, so that the bias in the coin flip is proportional to the difference in wealth of the two agents, normalized by the average wealth.  Note that if $w>x$ the coin is biased in favor of wealth going from Agent 2 to Agent 1, and vice versa.  If the two agents have equal wealth, the above reduces to $E[\eta]=0$ so the coin is unbiased.  We can then demand $E[\eta^2]=1$ without loss of generality.

The second term in Eq.~(\ref{eq:rw1}) implements a flat wealth tax with rate per unit time equal to $\chi$, which we call the {\it redistribution coefficient}.  The amount of wealth tax collected from Agent 1 in time $\Delta t$ is $(\chi\Delta t)w$, so that collected from the entire population in time $\Delta t$ is $(\chi\Delta t)W$.  The latter quantity is divided by $N$ and redistributed, so the net gain of Agent 1 is precisely the second term in Eq.~(\ref{eq:rw1}).  Note that agents with wealth less than the mean $W/N$ receive net benefit from the redistribution, at the expense of those above the mean.  The term is identical to that in the Fokker-Planck equation for the Ornstein-Uhlenbeck process~\cite{bib:OU}, and has the same stabilizing effect.

A Fokker-Planck equation~\cite{bib:Fokker, bib:Planck, bib:Kolmogorov} for the agent density distribution $P(w,t)$ of the above-described random process is found by letting $\Delta t \to 0$, which is the so-called {\it small-transaction limit}, and by supposing that the wealth of Agent 2, namely $x$, is also a random variable.  We define the {\it drift coefficient} and the {\it diffusivity} by
\begin{eqnarray}
\sigma &=& \lim_{\Delta t \to 0} \calE\left[\frac{\Delta w}{\Delta t}\right]
\label{eq:sigma}\\
\noalign{\noindent{\mbox{and}}}\nonumber\\
D &=& \lim_{\Delta t \to 0} \calE\left[\frac{(\Delta w)^2}{\Delta t}\right],
\label{eq:D}
\end{eqnarray}
respectively, where the expected value $\calE[f]$ of a function $f(\eta, x)$ over the distribution $P(x, t)$ is given by
\begin{equation}
\calE[f] = \frac{1}{N} \int_0^\infty \diff x \ P(x,t) E[f(\eta, x)].
\label{eq:exp}
\end{equation}
In terms of $\sigma$ and $D$, the Fokker-Planck equation can be written as:
\begin{eqnarray}
\frac{\partial P}{\partial t}
=
-\frac{\partial}{\partial w} \left(\sigma P\right)
+ \frac{1}{2} \frac{\partial^2}{\partial w^2} \left(DP\right).
\label{eq:fpgen}
\end{eqnarray}

The explicit calculation of $\sigma$ and $D$ follows straightforwardly from Eqs.~(\ref{eq:rw1}), (\ref{eq:rw2}), (\ref{eq:sigma}), (\ref{eq:D}) and (\ref{eq:exp}); details are given in a reference~\cite{bib:Boghosian2017}.  The final result is the nonlinear, partial integrodifferential Fokker-Planck equation for the EYSM,
\begin{eqnarray}
\lefteqn{
\frac{\partial P}{\partial t}
=
-\frac{\partial}{\partial w}\left[\chi\left(\frac{W}{N}-w\right)P\right]}\nonumber\\
& &
+\frac{\partial}{\partial w}
\left\{\zeta\left[2\frac{N}{W}\left(B-\frac{w^2}{2}A\right) + \left(1-2L\right)w\right]P\right\}\nonumber\\
& &
+\frac{\partial^2}{\partial w^2}\left[\gamma\left(B + \frac{w^2}{2}A\right)P\right],
\label{eq:fp}
\end{eqnarray}
where we have defined the {\it Pareto-Lorenz potentials},
\begin{eqnarray}
A(w,t) &:=& \frac{1}{N}\int_w^\infty dx\; P(x,t)
\label{eq:A}\\
L(w,t) &:=& \frac{1}{W}\int_0^w dx\; P(x,t)x
\label{eq:L}\\
\noalign{\noindent{\mbox{and}}}\nonumber\\
B(w,t) &:=& \frac{1}{N}\int_0^w dx\; P(x,t)\frac{x^2}{2},
\label{eq:B}
\end{eqnarray}
and where the total number of agents and the total wealth are given in terms of $P(w,t)$ by
\begin{eqnarray}
N &:=& \int_0^\infty dw\; P(w,t)
\label{eq:N}\\
\noalign{\noindent{\mbox{and}}}\nonumber\\
W &:=& \int_0^\infty dw\; P(w,t)w
\label{eq:W},
\end{eqnarray}
respectively.  It is straightforward to show that $N$ and $W$ are constants of the motion of Eq.~(\ref{eq:fp}).

We can divide Eq.~(\ref{eq:fp}) by $\gamma$, absorbing it into the time $t$, as well as into the redistribution and WAA coefficients, $\chi$ and $\zeta$, respectively.  Equivalently stated, by setting $\gamma$ to unity we obtain the form of the equation for natural {\it transactional} units of time $t$, and we can think of $\chi$ and $\zeta$ as coefficients for redistribution and WAA on a {\it per transaction} basis, respectively.  Hence, with these transactional units adopted, the EYSM is seen to have only two free parameters.

In steady-state, we set the time derivative to zero in Eq.~(\ref{eq:fp}), and integrate once with respect to $w$ to obtain the nonlinear, ordinary integrodifferential equation
\begin{eqnarray}
\lefteqn{
\frac{\diff}{\diff w}\left[\left(B + \frac{w^2}{2}A\right)P\right] = \chi \left(\frac{W}{N}-w \right) P}
\;\;\;\;
\nonumber \\
& &
- 2 \zeta \left[ \frac{N}{W}\left(B-\frac{w^2}{2}A\right) + \left(\frac{1}{2}-L\right)w \right]P.
\label{eq:fpss}
\end{eqnarray}
Henceforth, we focus on the classical and/or weak solutions of this steady-state equation, and so we ignore the time dependence, writing $P(w)$ instead of $P(w,t)$, etc.

\subsection{Lorenz curve and Gini coefficient}
Though the agent density function $P(w)$ is very fundamental, an alternative representation called the {\it Lorenz curve}~\cite{bib:Lorenz} is widely used for analyzing wealth distributions.  If we know the agent density function $P(w)$, the Lorenz curve can be obtained by plotting the fraction of total wealth held by agents with wealth less than $w$, namely $L(w) := \frac{1}{W} \int_0^w dx\; P(x) x$, versus the fraction of agents with wealth less than $w$, namely $F(w) := \frac{1}{N}\int_0^w dx\; P(x)$, where $N$ and $W$ are given by Eqs.~(\ref{eq:N}) and (\ref{eq:W}).  This is a curve in the unit square, parametrized by the wealth $w$, that can be shown to be concave up and lie below the diagonal.  Since $F(w)$ increases monotonically from 0 to 1, it can be inverted to obtain $w=F^{-1}(f)$ for $f$ on that interval.  The functional form of the Lorenz curve is then denoted by $\calL(f) := L(F^{-1}(f))$ for $f\in[0,1]$.

\begin{figure*}
        \begin{subfigure}[b]{0.45\textwidth}
            \begin{center}
            \includegraphics[width=\textwidth]{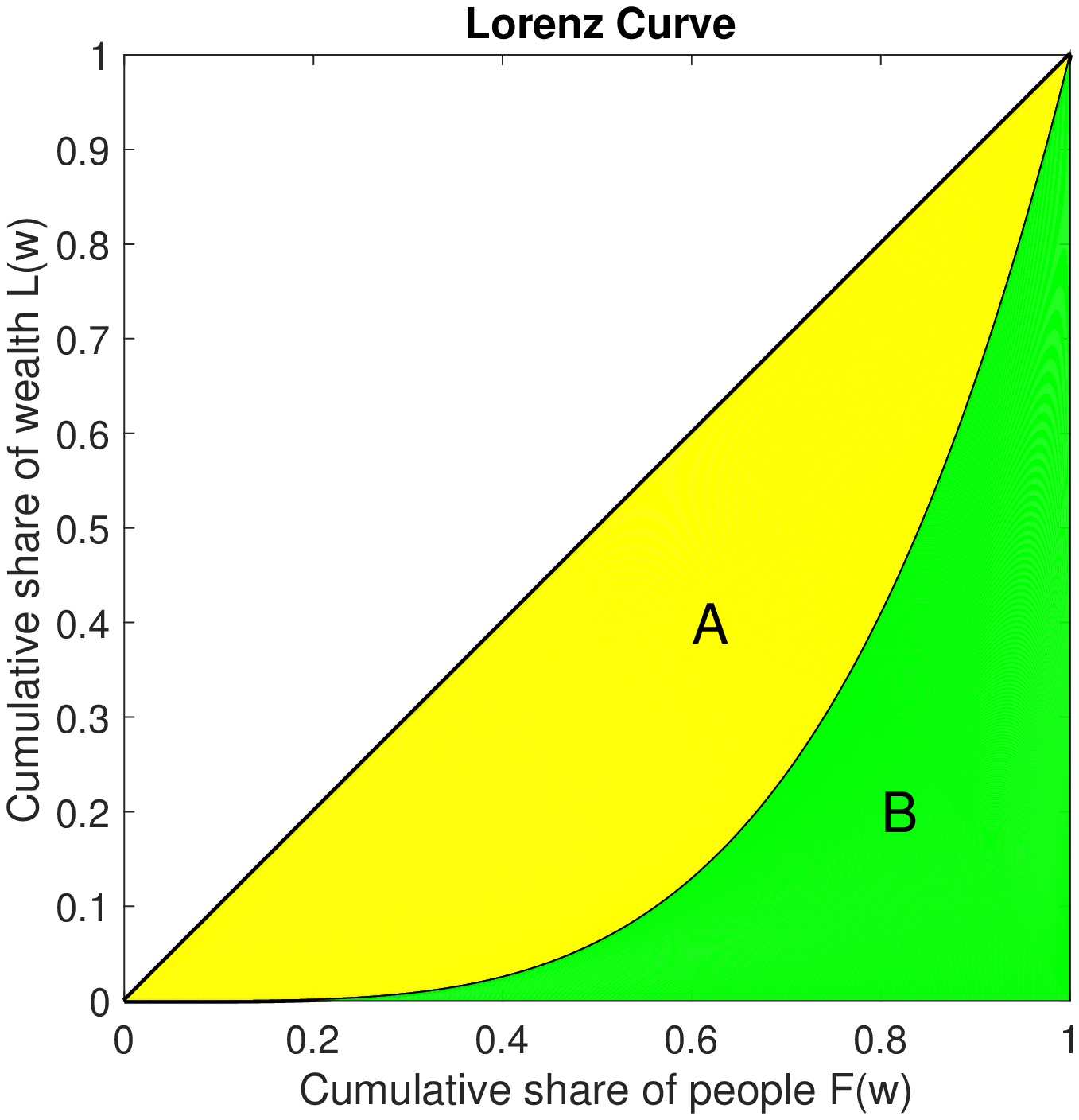}
            \caption{{\small Lorenz Curve for subcritical state with classical wealth density function}}
            \label{fig:subcriticalLorenzEYSM}
            \end{center}
        \end{subfigure}
        \begin{subfigure}[b]{0.45\textwidth}
            \begin{center}
            \includegraphics[width=\textwidth]{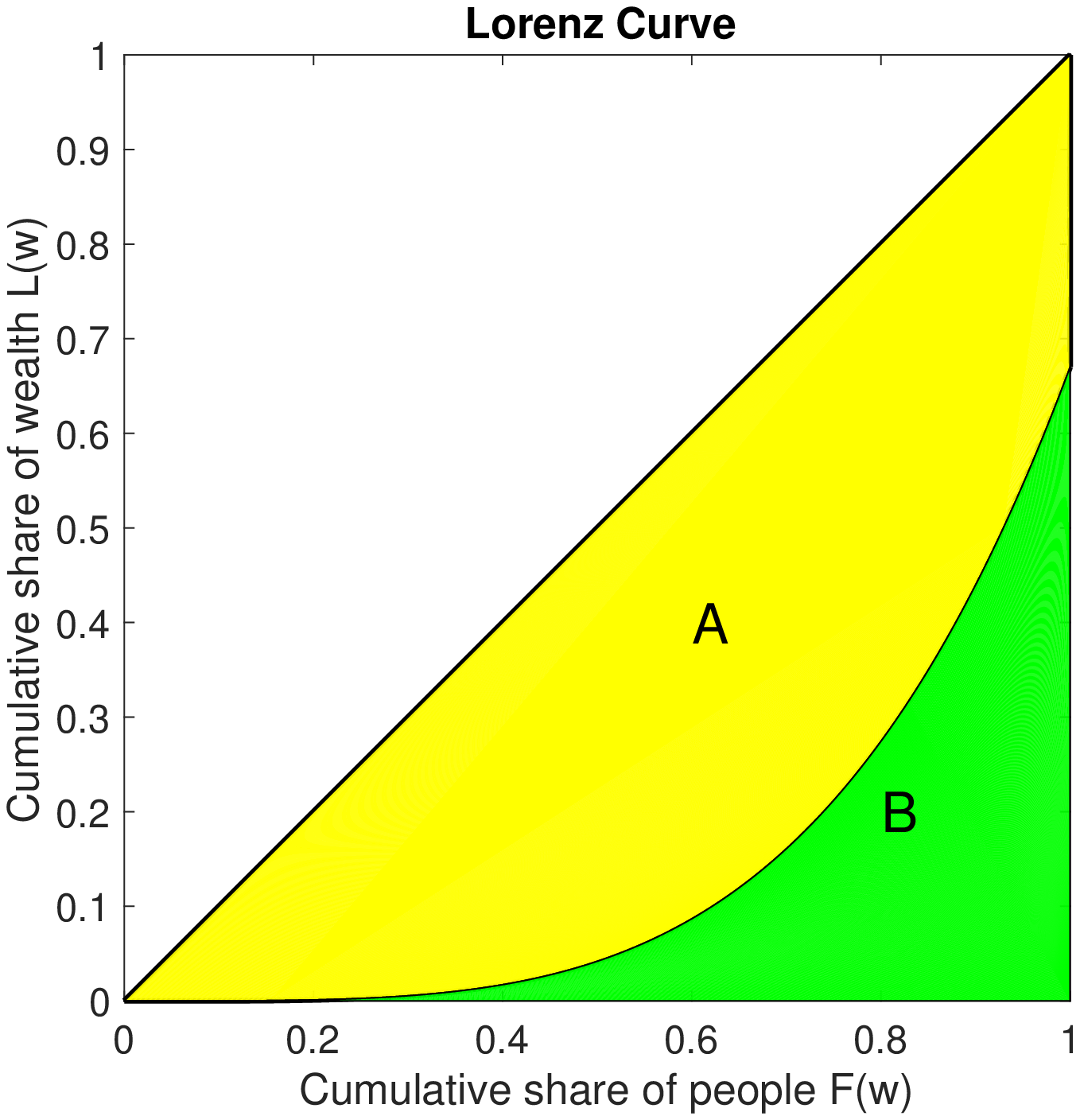}
            \caption{{\small Lorenz Curve for supercritical state with partial wealth condensation}}
            \label{fig:supercriticalLorenzEYSM}
            \end{center}
        \end{subfigure}
        \caption{{\small In the subcritical case, the Lorenz curve cannot go below zero and always terminates at point $(1,1)$.  The Gini coefficient is defined to be $G = \frac{A}{A + B}$, where $A$ and $B$ are the areas of the shaded regions.}}
\end{figure*}

Since zero percent of the agents have zero percent of the wealth, and one hundred percent of the agents have one hundred percent of the wealth, one might reasonably expect that $\calL(0)=0$ and $\calL(1)=1$, as in Fig.~\ref{fig:subcriticalLorenzEYSM}, and, indeed, this is exactly what does happen when $\zeta\leq\chi$, which we call the {\it subcritical} regime.  When $\zeta > \chi$, however, it has been shown~\cite{bib:Boghosian2017} that $\calL(1) = \sfrac{\chi}{\zeta} < 1$, indicative of partial wealth condensation, as in Fig.~\ref{fig:supercriticalLorenzEYSM}.  In essence, a fraction of $1 - \sfrac{\chi}{\zeta}$ of the total wealth is held by a vanishingly small number of agents, which we refer to as the {\it oligarchy}.  Mathematically, this phenomenon is best understood using the methods of distribution theory or of nonstandard analysis~\cite{bib:DevittLeeWangLiBoghosian}, but we shall eschew those topics in this presentation.

Since $F'(w) = P(w)/N$ and $L'(w) = P(w)w/W$, it follows that the slope of the Lorenz curve,
\begin{equation}
\calL'(f) = \left.\frac{\;\frac{dL}{dw}\;}{\;\frac{dF}{dw}\;}\right|_{w = F^{-1}(f)} = \frac{F^{-1}(f)}{W/N},
\end{equation}
is just the wealth parameter divided by the average wealth.  Hence, as long as all agents have positive wealth, which is always the case in the EYSM, the Lorenz curve will be monotonically increasing.  Later, in Sec.~\ref{sec:AWM} of this paper, we will consider a model that allows for negative-wealth agents, for which the Lorenz curve may first decrease from the point $(0,0)$ before eventually turning upward.

Given the Lorenz curve as well as $N$ and $W$, the distribution $P(w)$ can be recovered.  To see how this is done, first note that if $\calL(f)$ is known, the derivative $\frac{d\calL}{df}$ can be calculated as a function of $f$.  Since $N$ and $W$ are additionally known, it follows that the agent wealth
\begin{equation}
w = \frac{W}{N}\frac{d\calL}{df}
\end{equation}
is known as a function of $f$.  Because $\calL(f)$ is concave up, $w$ will be a monotonically increasing function of $f$, so it can be inverted to obtain $f=F(w)$ as a function of $w$, which may finally be differentiated and multiplied by $N$ to obtain $P(w)=N F'(w)$.  Since we have already noted that $\{\calL(f), N, W\}$ can be computed from $P(w)$, and since we have just shown that the opposite is also true, it follows that $\{\calL(f), N, W\}$ and $P(w)$ contain equivalent information.

For a hypothetical society in which all agents have exactly the same wealth, it is straightforward to see that $L$ and $F$ would be equal, so the Lorenz curve would be a diagonal line.  An important measure of wealth inequality is then the {\it Gini coefficient} $G$~\cite{bib:Gini1, bib:Gini2}, defined to be twice the area between the Lorenz curve and the diagonal.  In the above-described case of complete economic equality, we would have $G=0$.  In a complete oligarchy, we would have the degenerate Lorenz curve $\calL(f)=0$ for all $0\leq f<1$, so the area between the Lorenz curve and the diagonal would be $\sfrac{1}{2}$, and hence $G=1$.  In the more realistic situation for which the Lorenz curve lies between these extremes, $0<G<1$.  (With a sufficiently large number of negative-wealth agents, however, it is actually possible to have $G>1$.)  It is straightforward to show that
\begin{equation}
G = 1 - \frac{2}{W}\int_0^\infty dw\; P(w)A(w)w.
\label{eq:Gini}
\end{equation}

In the remainder of this paper, instead of trying to fit model agent density functions to their empirical counterparts, we will focus on fitting Lorenz curves.  This is more appropriate for a number of reasons, including
\begin{itemize}
\item Lorenz curves and their associated Gini coefficients are widely used and recognized metrics for analyzing wealth distributions.
\item Empirical agent density functions have non-compact support in the continuum limit, whereas the Lorenz curve $\calL(f)$ is known to be concave up and, conveniently, supported on $f\in[0,1]$.
\item The supercritical regime, in which wealth is condensed into a partial or complete oligarchy, is not easily detectable from the agent density function, but it is immediately obvious from inspection of the Lorenz curve.  The value of $\lim_{f\rightarrow 1^-}\calL(f)$, where the Lorenz curve intersects the right boundary, is the fraction of wealth held by the non-oligarchical part of the population.
\item Finally, as has been demonstrated, Lorenz curves contain as much information as agent density functions that have been scaled so that $N=W=1$.  In fact, this scaling is a desirable normalization when comparing wealth distributions in different countries that may have very different values of $N$ and $W$.
\end{itemize}

\subsection{Symmetries and Duality}
\label{ssec:SymDual}

Eq.~(\ref{eq:fpss}) exhibits two remarkable symmetries.  We might anticipate the first of these from our discussion of the Lorenz curve above.  Let's say that $P(w;\chi,\zeta;W,N)$ denotes the solution to Eq.~(\ref{eq:fpss}) for redistribution coefficient $\chi$, WAA coefficient $\zeta$, total number of agents $N$, and total amount of wealth $W$.  Let's suppose that we know the Lorenz curve for this wealth distribution, but we do not know $N$ and $W$.  In that case, we could make the formal assumption that $N=W=1$ and thereby find $P(w;\chi,\zeta;1,1)$.  Then it is readily verified that the solution to Eq.~(\ref{eq:fpss}) for other values of $N$ and $W$ is given by
\begin{equation}
P(w;\chi,\zeta;N,W) = \frac{N}{W/N} P\left(\frac{w}{W/N};\chi,\zeta;1,1\right).
\label{eq:symm1}
\end{equation}
Hence, if Eq.~(\ref{eq:fpss}) can be solved formally for $N=W=1$, the graph of the solution can be scaled both horizontally and vertically to obtain the solution for any $N$ and $W$ whatsoever.

The second symmetry is not at all obvious.  It was derived in a recent reference~\cite{bib:LiBoghosian2017}, both from the steady-state Fokker-Planck equation, Eq.~(\ref{eq:fpss}), and from the underlying statistical process.  Let
\begin{equation}
\calL^{\mbox{\tiny sup}}_{\langle\chi,\zeta\rangle}(f) := L(F^{-1}(f;\chi,\zeta);\chi,\zeta)
\end{equation}
denote a supercritical Lorenz curve function with redistribution coefficient $\chi$ and WAA coefficient $\zeta > \chi$.  It follows that swapping $\chi$ and $\zeta$ will result in a subcritical Lorenz curve function $\calL^{\mbox{\tiny sub}}_{\langle\chi,\zeta\rangle}(f)$.  Then it has been shown that
\begin{equation}
\calL^{\mbox{\tiny sup}}_{\langle\chi,\zeta\rangle}(f)
=
\frac{\chi}{\zeta}\calL^{\mbox{\tiny sub}}_{\langle\zeta,\chi\rangle}(f),
\label{eq:symm2}
\end{equation}
for $f\in[0,1)$, still assuming that $\zeta > \chi$.

This establishes a one-to-one correspondence between the subcritical and supercritical states of the system.  This symmetry is an example of the phenomenon of {\it duality}, well known in physics.  In our application, duality turns out to be very useful for numerical computation of the solutions, by eliminating the need to compute supercritical solutions.  If you need to compute a supercritical Lorenz curve, you can simply compute the subcritical one obtained by swapping $\chi$ and $\zeta$, and then rescale in the manner described above.  A more detailed description of duality can be found in one of the references~\cite{bib:LiBoghosian2017}.

\section{The Affine Wealth Model}
\label{sec:AWM}
\subsection{Definition of model}
\label{ssec:AWMDef}

A key deficiency of the EYSM is its inability to account for negative wealth.  Agent wealth is initially positive, and the dynamics keep it so by design, so the agent density function $P$ is supported on $(0,\infty)$.  As noted earlier, however, roughly 10.9\% of the population of the United States in 2016 had negative wealth. To overcome this deficiency, we generalize the EYSM by demanding a new kind of symmetry.  In addition to the multiplicative scalings described by Eqs.~(\ref{eq:symm1}) and (\ref{eq:symm2}), we demand invariance under a certain additive shift of the wealth distribution, to be described below.  Because the new model is invariant under both scalings and shifts, we refer to it as the {\it Affine Wealth Model} (AWM).

In the AWM, the agent density function $P(w)$ is supported on $[-\Delta,\infty)$, where the fixed positive quantity $\Delta$ will be described shortly.  At the beginning of each transaction, the transacting agents both add $\Delta$ to their wealths.  With the positive wealths that result, they perform an EYSM transaction.  Finally, they both subtract $\Delta$ from their wealths to complete the transaction.  A nice feature of this approach is that it is unnecessary to modify the EYSM algorithm to deal with negative wealth.

Note that, even if both agents begin with positive wealths, $w$ and $x$, the quantity
\begin{equation}
\Delta w = \sqrt{\gamma\Delta t}\;\min (w+\Delta,x+\Delta)\eta
\end{equation}
may be larger than $w$ and/or $x$, so that an agent may lose more wealth than he/she currently possesses, and thereby end up with negative wealth.  The overall effect is to create an EYSM distribution in the ``shifted wealth,'' $\bar{w} = w + \Delta \in (0,\infty)$, but then to transform that distribution, $\bar{P}(\bar{w})$, to be one in terms of $w$ rather than $\bar{w}$, which is easily accomplished as follows,
\begin{equation}
P(w) = \bar{P}(\bar{w}) = \bar{P}(w + \Delta),
\label{eq:shift}
\end{equation}
where $\bar{P}$ is an EYSM distribution.

Now $dw = d\bar{w}$, so it follows that
\begin{eqnarray}
N &=& \int_{-\Delta}^\infty dw\; P(w)
    = \int_0^\infty d\bar{w}\; \bar{P}(\bar{w}) = \bar{N}\\
\noalign{\noindent{\mbox{and}}}\nonumber\\
W &=& \int_{-\Delta}^\infty dw\; P(w)w\nonumber\\
    &=& \int_0^\infty d\bar{w}\; \bar{P}(\bar{w}) (\bar{w}-\Delta) = \bar{W} - \Delta\bar{N},
\end{eqnarray}
and hence the average wealth, $\mu := W/N$ is given in terms of the shifted average wealth, $\bar{\mu} := \bar{W}/\bar{N}$, by
\begin{equation}
\mu = \bar{\mu} - \Delta.
\label{eq:muShift}
\end{equation}
Going forward, we write $\Delta$ as a fraction of the {\it shifted} average wealth, $\bar{\mu}$, which is guaranteed to be positive, $\Delta = \kappa\bar{\mu}$, where $\kappa\geq 0$ is a new parameter of the model.  It follows that $\mu = (1-\kappa)\bar{\mu}$, and hence
\begin{equation}
\Delta = \lambda\mu = \kappa\bar{\mu},
\label{eq:DeltaDef}
\end{equation}
where we have defined
\begin{equation}
\lambda := \frac{\kappa}{1-\kappa}
\label{eq:lambdaDef}
\end{equation}
for convenience. While there is nothing preventing $\kappa$ from exceeding one in principle, we shall see that empirically determined values of $\kappa$ tend to be very small.

In similar fashion, we can compute the Lorenz-Pareto potentials, $F$, $A$, $L$ and $B$, in terms of their barred counterparts as follows:
\begin{eqnarray}
F(w) &=& \bar{F}(\bar{w})\label{eq:FAWM}\\
A(w) &=& \bar{A}(\bar{w})\label{eq:AAWM}\\
L(w) &=& (1+\lambda)\bar{L}(\bar{w}) - \lambda \bar{F}(\bar{w})\label{eq:LAWM}\\
B(w) &=& \bar{B}(\bar{w}) - \kappa {\bar{\mu}}^2\left[ \bar{L}(\bar{w}) - \kappa\frac{\bar{F}(\bar{w})}{2}\right].\label{eq:BAWM}
\end{eqnarray}
The corresponding inverse transformations are then
\begin{eqnarray}
\bar{F}(\bar{w}) &=& F(w)\label{eq:FbarAWM}\\
\bar{A}(\bar{w}) &=& A(w)\label{eq:AbarAWM}\\
\bar{L}(\bar{w}) &=& (1-\kappa)L(w)+\kappa F(w)\label{eq:LbarAWM}\\
\bar{B}(\bar{w}) &=& B(w) + \lambda\mu^2\left[L(w) + \lambda\frac{F(w)}{2}\right].\label{eq:BbarAWM}
\end{eqnarray}
Note that the transformation reduces to the identity when $\kappa=\lambda=0$.

\subsection{Fokker-Planck equation for AWM}
\label{ssec:FPAWM}

With the above transformations in hand, and restoring the time dependence of $P$ for a moment, we wish to derive the Fokker-Planck equation for the AWM.  We do this by supposing that the shifted wealth distribution is the result of an EYSM, so that $\bar{P}$ and its associated barred Pareto-Lorenz potentials should obey a version of Eq.~(\ref{eq:fp}).

We begin by writing Eq.~(\ref{eq:rw1}) for the AWM,
\begin{equation}
\Delta \bar{w} = \sqrt{\gamma \Delta t}\; \min (\bar{w},\bar{x}) \eta
+ \chi \left( \bar{\mu} - \bar{w} \right) \Delta t.
\label{eq:rw1AWM}
\end{equation}
Note that
\begin{equation}
\bar{\mu} - \bar{w} = (\mu + \Delta) - (w + \Delta) = \mu - w,
\end{equation}
so that the redistribution term is invariant under the shift.

Next, we wish to modify Eq.~(\ref{eq:rw2}).  Clearly $\bar{w} - \bar{x} = w-x$, so the numerator of the fraction in parentheses in that equation is invariant.  In the denominator, we use $\bar{W}/\bar{N}$ since that is guaranteed to be positive.  The modified version of Eq.~(\ref{eq:rw2}) for the AWM is then
\begin{equation}
E[\eta] = \zeta \sqrt{\frac{\Delta t}{\gamma}} \left(\frac{\bar{w}-\bar{x}}{\bar{\mu}}\right).
\label{eq:rw2AWM}
\end{equation}

Since Eqs.~(\ref{eq:rw1AWM}) and (\ref{eq:rw2AWM}) are identical to Eqs.~(\ref{eq:rw1}) and (\ref{eq:rw2}) but for the presence of the bars, the equation obeyed by $\bar{P}$ is
\begin{eqnarray}
\lefteqn{
\frac{\partial\bar{P}}{\partial t}
+\frac{\partial}{\partial \bar{w}}\bigg\{
\chi\left(\bar{\mu}-\bar{w}\right)\bar{P}
}
\nonumber\\
& &
-\zeta\left[\frac{2}{\bar{\mu}}\left(\bar{B}-\frac{\bar{w}^2}{2}\bar{A}\right) + \left(1-2\bar{L}\right)\bar{w}\right]\bar{P}\bigg\}\nonumber\\
& &
\;\;\;\;\;
=
\frac{\partial^2}{\partial \bar{w}^2}\left[\gamma\left(\bar{B} + \frac{\bar{w}^2}{2}\bar{A}\right)\bar{P}\right].
\label{eq:fpAWMbar}
\end{eqnarray}
We now insert the transformation described in Eqs.~(\ref{eq:shift})--(\ref{eq:lambdaDef}), and Eqs.~(\ref{eq:FbarAWM})--(\ref{eq:BbarAWM}) of Subsection~\ref{ssec:AWMDef} to obtain, after some calculation, the Fokker-Planck equation for the AWM,
\begin{widetext}
\begin{eqnarray}
\frac{\partial P}{\partial t}
&+& \frac{\partial}{\partial w}
\left\{ \left(\chi - \kappa\zeta\right) \left( \mu - w \right) P
- (1-\kappa)\zeta\left[
\frac{2}{\mu}\left( B - \frac{w^2}{2} A \right) + \left( 1-2L \right)w
\right] P \right\}
\nonumber\\
& &
\;\;\;\;\;\;\;\;\;\;\;\;\;\;\;\;\;\;\;\;\;\;\;\;\;\;\;\;\;\;\;\;\;\;\;\;\;\;\;\;
=
\frac{\partial^2}{\partial w^2}\left\{ \gamma \left[
\left( B + \frac{w^2}{2} A \right) + \lambda\mu \left( \mu L + A w \right) + \frac{\lambda^2\mu^2}{2}
\right] P \right\}.
\label{eq:fpAWM}
\end{eqnarray}
\end{widetext}
where $w\in [-\lambda\mu,\infty)$.

The equation for the steady-state agent density function can be derived by setting $\frac{\partial P}{\partial t} = 0$ and integrating once with respect to $w$ to obtain
\begin{widetext}
\begin{eqnarray}
\lefteqn{
\frac{d}{dw}\left\{ \gamma \left[
\left( B + \frac{w^2}{2} A \right) + \lambda\mu \left( \mu L + A w \right) + \frac{\lambda^2\mu^2}{2}
\right] P \right\} }
\nonumber\\
& &
\;\;\;\;\;\;\;\;\;\;\;\;\;\;\;\;\;\;\;\;
= \left(\chi - \kappa\zeta\right) \left( \mu - w \right) P
- (1-\kappa)\zeta\left[
\frac{2}{\mu}\left( B - \frac{w^2}{2} A \right) + \left( 1-2L \right)w
\right] P,
\label{eq:fpssAWM}
\end{eqnarray}
\end{widetext}
again for $w\in [-\lambda\mu,\infty)$.

Eqs.~(\ref{eq:fpAWM}) and (\ref{eq:fpssAWM}) are one-parameter deformations of Eqs.~(\ref{eq:fp}) and (\ref{eq:fpss}), respectively.  The former reduce to the latter when $\kappa$ (and hence $\lambda=\frac{\kappa}{1-\kappa}$) is set to zero.  Though the deformed equations appear more complicated, they have the same basic structure -- namely a transaction term, a redistribution term and a WAA term.  Furthermore, the WAA term and the redistribution term of Eqs.~(\ref{eq:fpAWM}) and (\ref{eq:fpssAWM}) have the exact same form as those in Eqs.~(\ref{eq:fp}) and (\ref{eq:fpss}) but with two interesting changes: First, the WAA coefficient is scaled by $1-\kappa$; second, the redistribution rate $\chi$ is effectively reduced by $\kappa\zeta$. For the first observation, we need $\kappa < 1$ to keep the scaled WAA coefficient positive. From Eq.~(\ref{eq:DeltaDef}), this is also equivalent to having a positive mean wealth $\mu$. As will be shown later in Subsections~\ref{ssec:Comparison} and~\ref{ssec:Results}, empirical fittings suggest that reasonable values of $\kappa$ are all far less than one. The effect of a negative WAA coefficient on the solution to the differential equation is a mathematical question out of the scope of this paper. For the second observation, we need $\chi > \kappa \zeta$ to keep the reduced redistribution coefficient positive. As will be shown in Eq.~(\ref{eq:LorenzRightBoundary}), given that $\kappa < 1$, this is also equivalent to the Lorenz curve intersecting its right-hand boundary at a positive value, so that the wealth held by the non-oligarchical population is positive.  Again, empirical evidence presented later will show that this inequality is always satisfied by a large margin in reality, and we leave the case when it is violated as a mathematical question for future study. Our current intuition is that this may cause an instability or even non-existence of the solution to the differential equation.

\subsection{Lorenz curve and duality for the AWM}
\label{ssec:LorenzAWM}

To better understand the nature of the shift invariance, denote solutions to Eq.~(\ref{eq:fpssAWM}) by $P(w;\chi,\zeta,\kappa;W,N)$ for $w\in [-\lambda\mu,\infty)$.  From the construction of Eq.~(\ref{eq:fpssAWM}), it is clear that the shifted density function $\bar{P}$, obeys the Fokker-Planck equation for the EYSM, Eq.~(\ref{eq:fpAWMbar}), which is the same as that for the AWM when $\kappa=0$.  It follows that we have the new symmetry
\begin{equation}
P(w;\chi,\zeta,\kappa;W,N) = P(w+\lambda\mu;\chi,\zeta,0;(1+\lambda)W,N),
\label{eq:shiftAWM}
\end{equation}
and so it is possible to solve the Fokker-Planck equation for the AWM by solving the much simpler version for the EYSM and shifting the result.

The above observation gives us a complete strategy for solving for the agent density function for the AWM.  When asked to find $P(w;\chi,\zeta,\kappa;W,N)$ for $w\in [-\lambda\mu,\infty)$:
\begin{enumerate}
\item Use Eq.~(\ref{eq:shiftAWM}) to transform it to a problem for which $\kappa=0$ and $w\in[0,\infty)$.
\item If $\zeta > \chi$ so that the problem is supercritical, use Eq.~(\ref{eq:symm2}) to transform it to a subcritical one.
\item Finally, use Eq.~(\ref{eq:symm1}) to transform the problem to one in so-called {\it canonical form}, for which $N=W=1$.
\end{enumerate}
Hence, the only numerical solutions needed for solving the AWM are subcritical, canonical-form solutions for the EYSM.

We can go one step further and directly relate the Lorenz curves of the three stages of the above strategy.  We first consider the subcritical case, for which Step 2 is unnecessary.  Moreover, we can suppose that we start in canonical form, so that Step 3 is unnecessary.  We again denote the Lorenz curve of the subcritical AWM with redistribution coefficient $\chi$ and WAA coefficient $\zeta < \chi$ by $\calL^{\mbox{\tiny sub}}_{\langle\chi,\zeta\rangle}(f) = L(F^{-1}(f))$, and that of the corresponding EYSM solution by $\bar{\calL}^{\mbox{\tiny sub}}_{\langle\chi,\zeta\rangle}(f) = \bar{L}(\bar{F}^{-1}(f))$.  Then, using Eqs.~(\ref{eq:FAWM}) and (\ref{eq:LAWM}), we have
\begin{eqnarray}
\calL^{\mbox{\tiny sub}}_{\langle\chi,\zeta\rangle}(f)
&=& L(F^{-1}(f))\nonumber\\
&=& L(\bar{F}^{-1}(f))\nonumber\\
&=& (1+\lambda)\bar{L}(\bar{F}^{-1}(f)) - \lambda\bar{F}(\bar{F}^{-1}(f)),\nonumber
\end{eqnarray}
or
\begin{equation}
\calL^{\mbox{\tiny sub}}_{\langle\chi,\zeta\rangle}(f)
= (1+\lambda)\overline{\calL}^{\mbox{\tiny sub}}_{\langle\chi,\zeta\rangle}(f) - \lambda f,
\end{equation}
which directly relates the Lorenz curve of the subcritical AWM to that of the corresponding EYSM.  Note that
\begin{eqnarray}
\calL^{\mbox{\tiny sub}}_{\langle\chi,\zeta\rangle}(0)
&=&
(1+\lambda)0 - \lambda 0 = 0\\
\noalign{\noindent{\mbox{and}}}\nonumber\\
\calL^{\mbox{\tiny sub}}_{\langle\chi,\zeta\rangle}(1)
&=&
(1+\lambda)1 - \lambda 1 = 1,
\end{eqnarray}
as expected.

We next consider the supercritical case in which $\zeta > \chi$, again using the canonical form so that Step 3 is unnecessary.  A line of reasoning similar to that above yields
\begin{equation}
\calL^{\mbox{\tiny sup}}_{\langle\chi,\zeta\rangle}(f)
= (1+\lambda)\overline{\calL}^{\mbox{\tiny sup}}_{\langle\chi,\zeta\rangle}(f) - \lambda f.
\end{equation}
We can now use Eq.~(\ref{eq:symm2}) to rewrite this as
\begin{equation}
\calL^{\mbox{\tiny sup}}_{\langle\chi,\zeta\rangle}(f)
= (1+\lambda)\frac{\chi}{\zeta}\overline{\calL}^{\mbox{\tiny sub}}_{\langle\zeta,\chi\rangle}(f) - \lambda f,
\end{equation}
It still follows that
\begin{equation}
\calL^{\mbox{\tiny sup}}_{\langle\chi,\zeta\rangle}(0) = 0,
\end{equation}
but now we have
\begin{equation}
\calL^{\mbox{\tiny sup}}_{\langle\chi,\zeta\rangle}(1)
=
(1+\lambda)\frac{\chi}{\zeta} - \lambda
\label{eq:LorenzRightBoundary}
\end{equation}
for the fraction of wealth held by the non-oligarchical part of the population, and
\begin{equation}
1-\calL^{\mbox{\tiny sup}}_{\langle\chi,\zeta\rangle}(1)
=
(1+\lambda)\left(1-\frac{\chi}{\zeta}\right)
\label{eq:OligarchyWealthFraction}
\end{equation}
for the fraction of wealth held by the oligarchy.

Notice that, given $\kappa < 1$, $\calL^{\mbox{\tiny sup}}_{\langle\chi,\zeta\rangle}(1) > 0$ is equivalent to $\chi > \kappa \zeta$, and this is the condition we have discussed in subsection~\ref{ssec:FPAWM}.

\begin{figure*}
        \begin{subfigure}[b]{0.45\textwidth}
            \begin{center}
            \includegraphics[width=\textwidth]{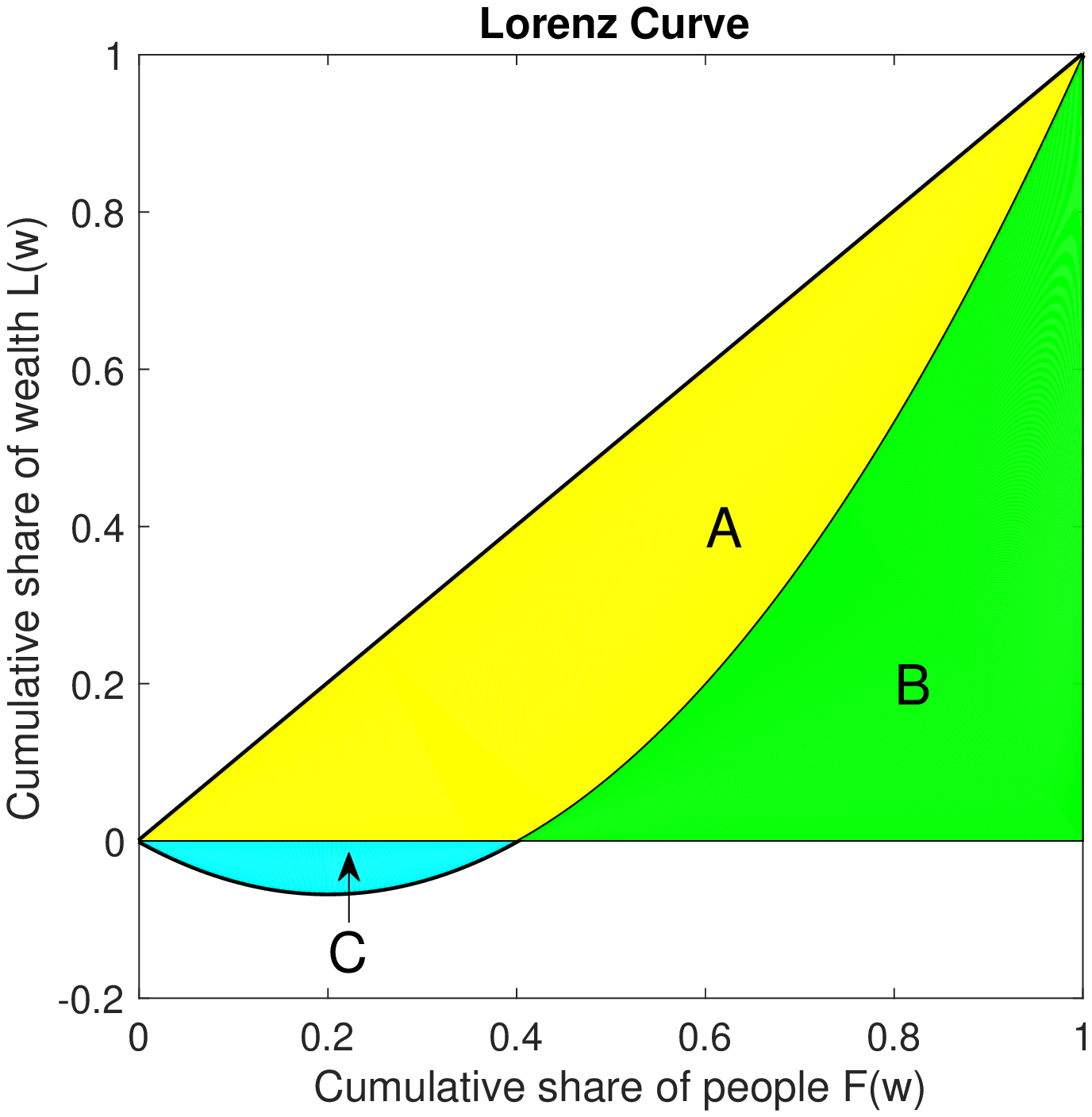}
            \caption{{\small Lorenz Curve for AWM subcritical state with negative wealth}}
            \label{fig:subcriticalLorenzAWM}
            \end{center}
        \end{subfigure}
        \begin{subfigure}[b]{0.45\textwidth}  
            \begin{center}
            \includegraphics[width=\textwidth]{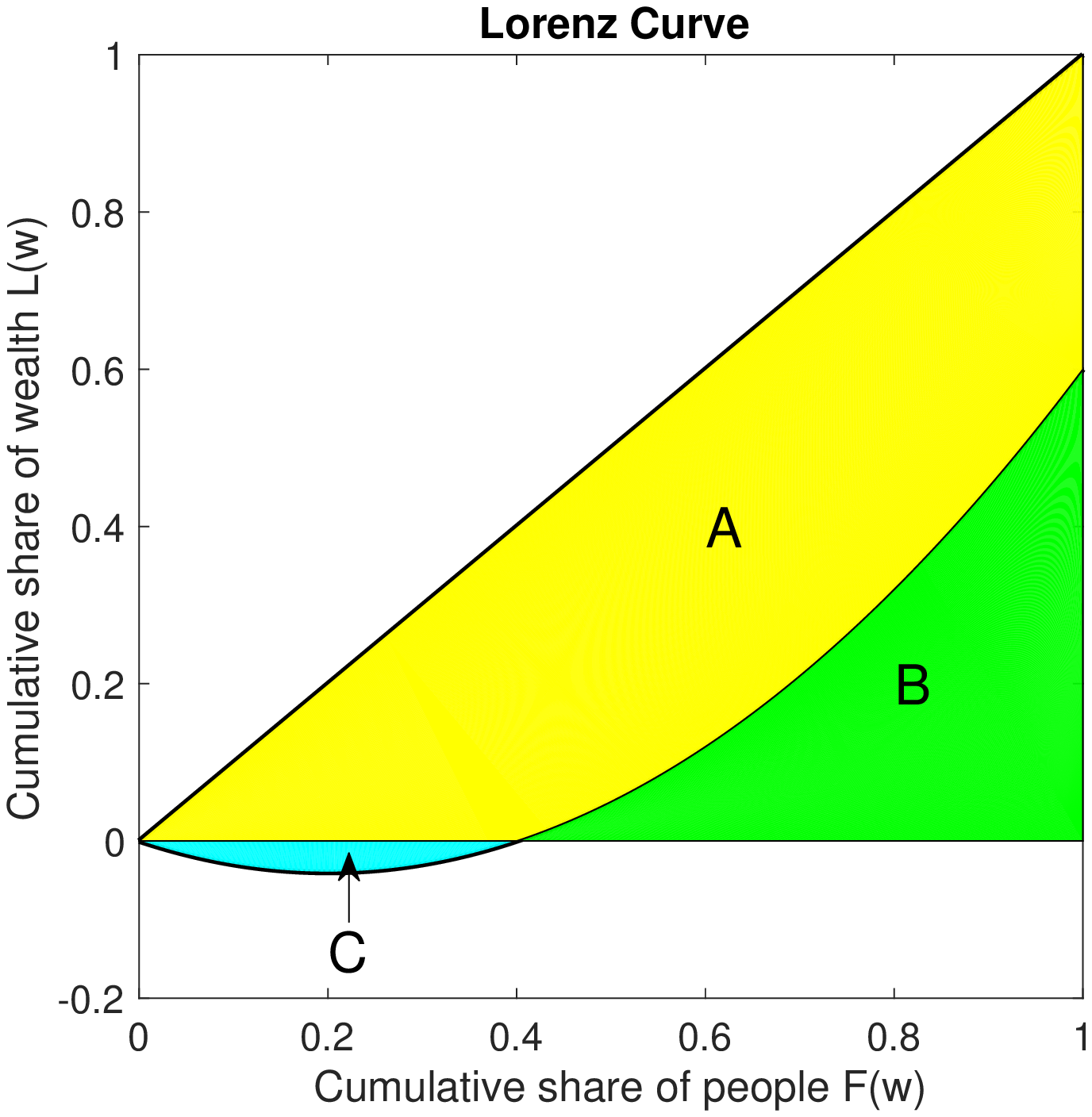}
            \caption{{\small Lorenz Curve for AWM supercritical state with negative wealth}}
            \label{fig:supercriticalLorenzAWM}
            \end{center}
        \end{subfigure}
        \caption{{\small In the subcritical case, the Lorenz curve dips below zero but terminates at point $(1,1)$.  The Gini coefficient is defined to be $G = \frac{A}{A + B}$, where $A$ and $B$ are the areas of the shaded region. With our AWM and in empirical wealth distributions, the Lorenz curve can: 1, go negative due to the existence of agents with negative total wealth; 2, hit the right boundary somewhere below point (1,1) due to the existence of oligarchy. The Gini coefficient in this case is define to be $G = \frac{A - C}{A + B - C}$, where $A$, $B$ and $C$ are the areas of the shaded regions.}}
\end{figure*}

Using the methodology described in this subsection, we can plot the Lorenz curve for the AWM for any given parameter triplet $\langle\chi,\zeta,\kappa\rangle$ by applying transformations to Lorenz curves for subcritical solutions of the EYSM.  This observation enormously facilitated obtaining the fitting results presented later in this paper.  Examples of subcritical and supercritical Lorenz curves for the AWM with negative-wealth agents are presented in Figs.~\ref{fig:subcriticalLorenzAWM} and \ref{fig:supercriticalLorenzAWM}.

\section{Empirical tests}
\subsection{Description of data used}
\label{ssec:Data}

The data we used for the U.S.\ wealth distribution was taken from the U.S.\ Survey of Consumer Finances (SCF) conducted by the Federal Reserve Board in cooperation with the U.S.\ Department of the Treasury~\cite{bib:SCF2013}.  It is a triennial cross-sectional survey of U.S.\ families, which includes information on families' balance sheets, pensions, income, and demographic characteristics.

Among the data fields collected for the households surveyed in the SCF is one called {\tt networth}, which represents the total wealth of a household~\footnote{Technically, {\tt networth} is not contained in the original microdata of the SCF.  Users of SCF data must calculate it themselves by summing over a number of other financial variables that are among the microdata.  In fact, an example explaining how to do exactly this is provided with the SCF data~\cite{bib:SCF2013}, and we followed this example closely when preparing the data for this study.}  Because {\tt networth} is calculated as the difference between assets and liabilities, its value can be negative.  Indeed, as noted earlier, about 10.9\% of the U.S.\ population has negative net wealth, and so the Lorenz curve for the U.S.\ actually does dip below zero, as described in Subsection~\ref{ssec:LorenzAWM}.

In the remainder of this section, we will compare several different models with empirical data from the SCF.  Of these models, only the AWM is capable of producing a Lorenz curve with negative values.  For the other models considered, there will necessarily be a significant error at low wealth, where the Lorenz curve of the model is positive, but that of the data is negative.

For reasons of confidentiality, the published SCF data intentionally excluded people listed on the so-called ``Forbes 400'' list of the wealthiest people in the U.S.~\cite{bib:Kennickell2001}.  Because this group of people, though small in number, are so wealthy as to have a nonnegligible impact of the overall distribution especially in the upper tail, we felt it important to add them back into the SCF data.  Fortunately, the journal {\it Forbes} publishes this list of people on an annual basis, including an estimate of their net wealth, so we simply merged the Forbes 400 dataset with the SCF dataset and used their union to conduct our analyses.  We checked the resulting dataset by using Eq.~(\ref{eq:Gini}) to calculate the wealth Gini coefficient, and comparing that to published values; for example, for the 2013 SCF data, we calculated a Gini coefficient of 85.5\%, which is very close to that was reported (85.1\%) in the ``Global Wealth Databook,'' published by Credit Suisse in 2013~\cite{bib:GlobalWealthDataBook}.

The empirical wealth distribution obtained in the manner described above is discretized by groups of households.  The $j$th such group is represented as having net wealth $w_j$, and weight $p_j$.  The weights $p_j$ are presumably proportional to the number of households in each group, and are normalized over the population so that $\sum_j p_j = 1$.  The density function of the wealth can therefore be written as a sum of weighted Dirac delta distributions,
\begin{equation}
P(w) = \sum_{j=1}^N p_j \delta(w-w_j).
\end{equation}
It is clear that $N = \int dw\; P(w) = \sum_j p_j = 1$, and $W = \int dw\; P(w) w = \sum_j p_j w_j$.  The $w_j$ can all be uniformly scaled so that this last quantity is also equal to one, so that the empirical data is in canonical form, with $N=W=1$.

To plot the Lorenz ordinates, we need to compute the cumulative sum of the population with wealth less than $w$ and their corresponding cumulative wealth.  This is equivalent to plotting the points $(f_j, \ell_j)$, where
\begin{eqnarray}
f_j &:=& \sum_{i\, :\, w_i \le w_j} p_i\\
\noalign{\noindent{\mbox{and}}}\nonumber\\
\ell_j &:=& \sum_{i\, :\, w_i \le w_j} p_i w_i.
\end{eqnarray}
The empirical Lorenz curve to which we compare our models in this paper is a linear interpolation of the Lorenz ordinates $(f_j, \ell_j)$, described above.  Since SCF data is very fine, including tens of thousands of households, the interpolation appears as a smooth curve, as does the numerical solution to our theoretical model.

\subsection{Fitting method}
\label{ssec:FittingMethod}

The fitting was done by minimizing the $L_1$ norm of the difference between the empirical Lorenz curve and the theoretical Lorenz curve obtained by numerical solution of our models.  Consistent with notation we have already adopted above, if a model's parameters are the components of a {\it parameter vector} $\theta$, we shall write $\calL_\theta (f)$ for the theoretical (model) Lorenz curve corresponding to that parameter vector.  If we then write $\calL(f)$ for the empirical Lorenz curve, our fitting methodology can be described mathematically as
\begin{eqnarray}
\theta_{\mbox{\tiny optimal}}
=
\argmin_{\theta} J(\theta),
\end{eqnarray}
where we have defined the {\it discrepancy}
\begin{eqnarray}
J(\theta) := \int_0^1 df\; \left| \calL(f) - \calL_\theta (f) \right|
\end{eqnarray}
The choice of the $L_1$ norm here is inspired by the definition of Gini coefficient; just as the Gini coefficient is half the area between the Lorenz curve and the diagonal, the discrepancy is equal to the area between the theoretical and empirical Lorenz curves.

The parameter vector $\theta$ will have different dimensions for different models.  In what follows, we shall consider models with between one and three parameters, so the dimension of $\theta$ ranges from one to three.  These models are described in detail in Subsection~\ref{ssec:ModelDescription}.  In all cases, there are no guarantees for the concavity of $J(\theta)$ and therefore we employ a global numerical search for the optimal parameter(s).

\begin{figure}[!h]
  %\vspace{0.4cm}
  \begin{center}
  {\epsfig{file = 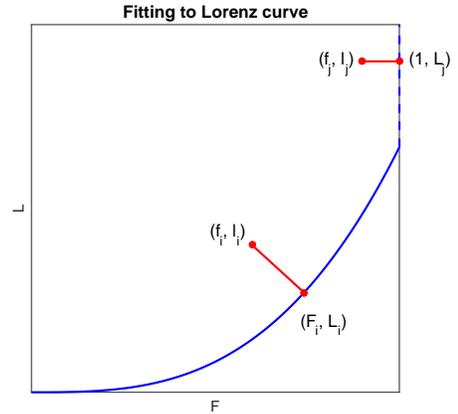, width = 8.0cm}}
  \end{center}
  \caption{Geometry of computing the local errors for the data points $(f_i,\ell_i)$ and $(f_j,\ell_j)$ to the fitting curve $\calL(f)$. For $(f_i, \ell_i)$, we compute its perpendicular distance to the point $(f,\calL_\theta(f))$, which is the closest point on $\calL_\theta(f)$. While for $(f_j, \ell_j)$, we compute its horizontal distance to the boundary $f = 1$.}
  \label{fig:fitting}
  \vspace{-0.3cm}
\end{figure}

For some of the results presented in Subsection~\ref{ssec:Results}, the model and empirical Lorenz curves are so close that it is difficult to distinguish them graphically.  For this reason, we display the local error between the two curves, plotted versus $F$, in an inset to each of the figures presented.  Since the slope of the Lorenz curve ranges from zero (or slightly less) to infinity, defining the local error as the vertical distance between the two curves would be misleading.  Instead, we define the local error as the length of a line segment connecting the empirical data point $(f_j,\ell_j)$ to the model Lorenz curve, constructed so as to be perpendicular to the latter, as shown in Fig.~\ref{fig:fitting}.  If there is more than one such line segment, the length of the shortest is used.  In other words, the local error is the shortest distance from the empirical data point to the model Lorenz curve.

For model Lorenz curves in the supercritical regime, when $\calL$ is greater than the fraction of wealth held by the non-oligarchical population and less than one, the solution will coincide with the boundary $f = 1$.  A line segment perpendicular to this part of the model Lorenz curve is, therefore, horizontal, so the local error is the horizontal distance from the point $(f_j,\ell_j)$ to the vertical boundary $f = 1$, i.e., it is equal to $| f_j - 1|$.  This is also shown in Fig.~\ref{fig:fitting}.

\subsection{Description of models used}
\label{ssec:ModelDescription}

\subsubsection{Single-agent model}

As a baseline for our comparisons, we use a linear model similar to one employed in earlier work by a number of authors.  (See, e.g., \cite{bib:BouchaudMezard2000}).  Instead of randomly selecting pairs of agents to engage in binary transactions, the model selects single agents to engage in unary transactions.  For this reason, we henceforth refer to it as the Single-Agent Model (SAM).  In a transaction, an agent with wealth $w$ has even odds of winning or losing a fraction $\sqrt{\gamma\Delta t}$ of his/her own wealth.  If we again employ a redistribution term of Ornstein-Uhlenbeck form, as in Eq.~(\ref{eq:rw1}), the analog of that equation is
\begin{eqnarray}
\Delta w &=& \sqrt{\gamma \Delta t}\; w \eta + \chi \left( \frac{W}{N} - w \right)\Delta t.
\end{eqnarray}
where $E[\eta]=0$ and $E[\eta^2]=1$.  This statistical process obviously conserves the total number of agents, but it conserves wealth only in a mean sense.  The easily derived linear Fokker-Planck equation corresponding to this model,
\begin{eqnarray}
\frac{\partial P(w,t)}{\partial t} &=& - \frac{\partial }{\partial w} \left[ \chi \left( \frac{W}{N} - w \right) P(w,t) \right] \nonumber \\
& & +\frac{\partial^2 }{\partial w^2} \left[ \gamma \frac{w^2}{2} P(w,t) \right],
\label{eq:fpSAM}
\end{eqnarray}
however, conserves both $N$ and $W$.  In the same spirit as our derivations of Eqs.~(\ref{eq:fpss}) and (\ref{eq:fpssAWM}), we see that the steady-state solutions of Eq.~(\ref{eq:fpSAM}) are described by
\begin{eqnarray}
\frac{\diff }{\diff w} \left[ \frac{w^2}{2} P(w) \right] = \chi \left(\frac{W}{N} - w \right) P(w),
\label{eq:fpSAMss}
\end{eqnarray}
which can be solved analytically.  If the constant of integration is chosen to satisfy the normalization requirement of Eq.~(\ref{eq:N}), and if we adopt transactional units by taking $\gamma=1$ as before, the result for the agent density function is
\begin{eqnarray}
P(w)
&=&
\frac{N}{\mu}
\frac{(2\chi)^{2\chi}}{\Gamma(2\chi)}
\left(\frac{\mu}{w}\right)^{2\chi+2}e^{-2\chi\frac{\mu}{w}},
\label{eq:PSAM}
\end{eqnarray}
where $\mu:=W/N$ is the mean wealth.  Notice that near $w=0$ this solution for $P(w)$ is very flat and depleted, while for $w$ very large, it is asymptotically a power-law consistent with the observations of Pareto.

From Eq.~(\ref{eq:PSAM}), it is straightforward to calculate
\begin{eqnarray}
F(w) &=& Q\left(2\chi+1,\frac{2\chi\mu}{w}\right)\\
\noalign{\noindent{\mbox{and}}}\nonumber\\
L(w) &=& Q\left(2\chi,\frac{2\chi\mu}{w}\right),
\end{eqnarray}
where $Q := \frac{\Gamma(a,z)}{\Gamma(a)}$ is the regularized upper incomplete gamma function.  The inverse of this function is typically denoted by $Q^{-1}(a,z)$, so that $Q(a,Q^{-1}(a,z))=z$, in terms of which the one-parameter Lorenz curve function is
\begin{equation}
\calL^{\mbox{\tiny SAM}}_{\langle\chi\rangle}(f)
=
Q\left(2\chi,Q^{-1}\left(2\chi+1,f\right)\right)
\end{equation}
Note that the parameter vector for this model, $\theta=\langle\chi\rangle$, is one-dimensional since the Lorenz curve depends only on the redistribution coeffiient $\chi$.

\subsubsection{EYSM with redistribution}

The second model that we consider is the EYSM as described in Subsection~\ref{ssec:EYSM}, with redistribution coefficient $\chi$, but without WAA so $\zeta=0$.  Again, the parameter vector is one-dimensional, and we denote the functional form of the Lorenz curve by $\calL^{\mbox{\tiny EYSM}}_{\langle\chi\rangle}(f)$.

\subsubsection{EYSM with redistribution \& WAA}

The third model that we consider is the EYSM as described in Subsection~\ref{ssec:EYSM}, but this time with both redistribution coefficient $\chi$, and WAA coefficient $\zeta$.  The parameter vector is therefore two-dimensional, and we denote the functional form of the Lorenz curve by $\calL^{\mbox{\tiny EYSM}}_{\langle\chi,\zeta\rangle}(f)$.

\subsubsection{AWM}

The fourth model that we consider is the AWM as described in Sec.~\ref{sec:AWM}.  The parameter vector for that model is three-dimensional, $\theta = \langle\chi,\zeta,\kappa\rangle$, and from the discussion in Subsection~\ref{ssec:LorenzAWM}, we know that we can write
\begin{equation}
\calL^{\mbox{\tiny AWM}}_{\langle\chi,\zeta,\kappa\rangle}(f)
=
(1+\lambda)\calL^{\mbox{\tiny EYSM}}_{\chi, \zeta}(f) - \lambda f,
\label{eq:LAWMEYSM}
\end{equation}
where it should be recalled that $\lambda$ is given by Eq.~(\ref{eq:lambdaDef}).

Now a global search in a three-dimensional parameter space would be computationally expensive.  Notice, however, that if we were using the $L_2$ norm to define the discrepancy instead of the $L_1$ norm, and if $\chi$ and $\zeta$ were held fixed, the optimal value of $\lambda$ would be given by
\begin{eqnarray}
\lambda^{L_2}_{\text{opt}}
=
\frac{
\int_0^1 df\;
\left[\calL^{\mbox{\tiny EYSM}}_{\langle\chi,\zeta\rangle}(f) - \ell(f)\right]
\left[f - \calL^{\mbox{\tiny EYSM}}_{\langle\chi,\zeta\rangle}(f)\right]}
{\int_0^1 df\; \left[f - \calL^{\mbox{\tiny EYSM}}_{\langle\chi,\zeta\rangle}(f)\right]^2},
\nonumber\\
\end{eqnarray}
where $\ell(f)$ is the empirical Lorenz curve.  From this we could compute
\begin{equation}
\kappa^{L_2}_\text{opt} = \frac{\lambda^{L_2}_\text{opt}}{1+ \lambda^{L_2}_\text{opt}}.
\end{equation}
In our numerical simulations, we used $\kappa^{L_2}_\text{opt}$ as the initial guess in a line search for the true optimal value $\kappa^{L_1}_\text{opt}$, obtained by minimizing the $L_1$ norm of the discrepancy.

In the above-described fashion, a three-dimensional optimization problem is reduced to a two-dimensional one in $\langle\chi,\zeta\rangle$ -- albeit with a quick line search at each point, for which we have an excellent initial guess.  We have found this method to work reliably and significantly reduce the computational work involved.

\subsection{Comparison of models}
\label{ssec:Comparison}

\begin{figure*}
	\centering
        \begin{subfigure}[b]{0.475\textwidth}
            \centering
            \includegraphics[width=\textwidth]{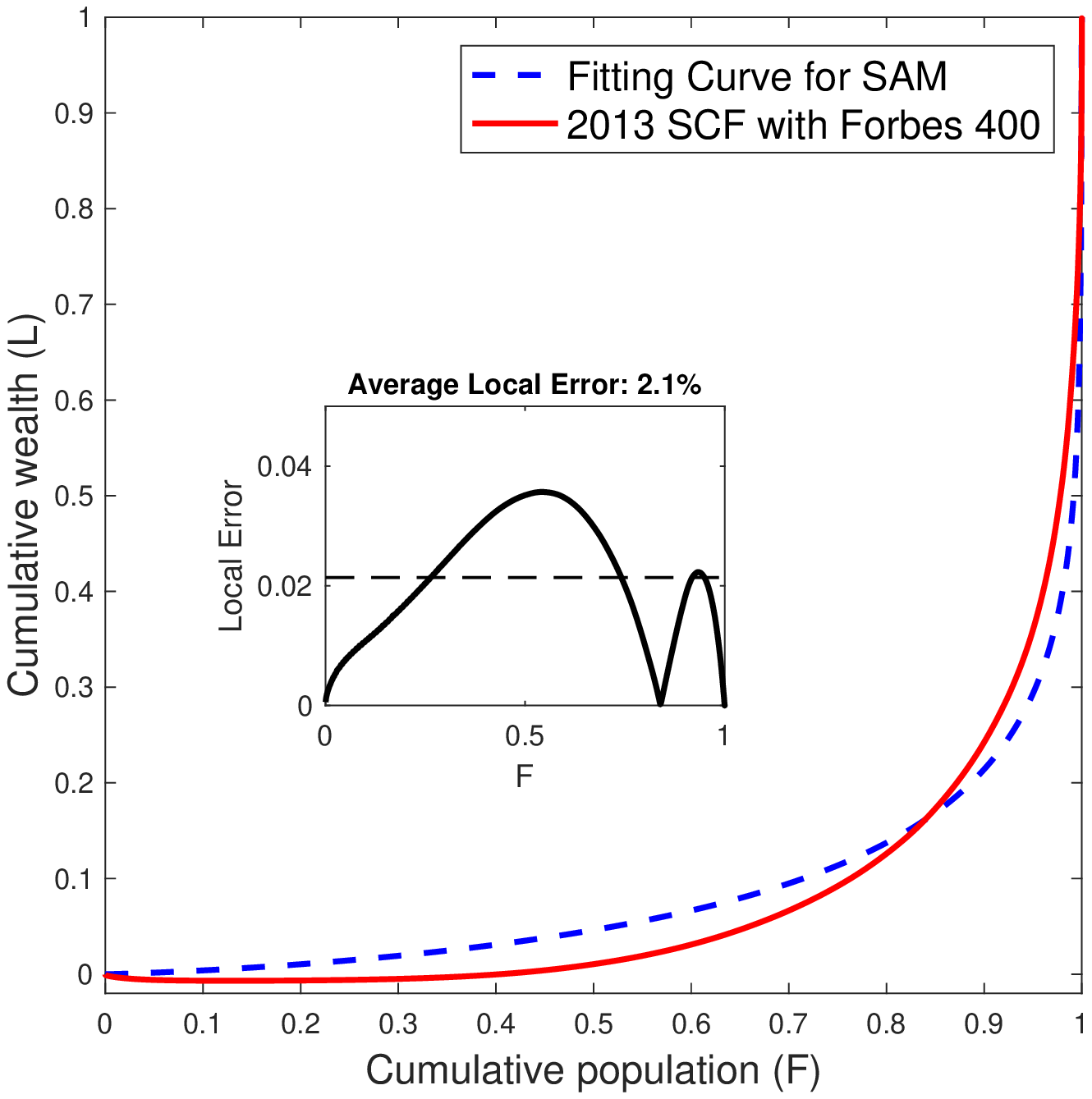}
            \caption{{\small Single Agent Model}}    
            \label{fig:comp1}
        \end{subfigure}
        \hfill
        \begin{subfigure}[b]{0.475\textwidth}  
            \centering 
            \includegraphics[width=\textwidth]{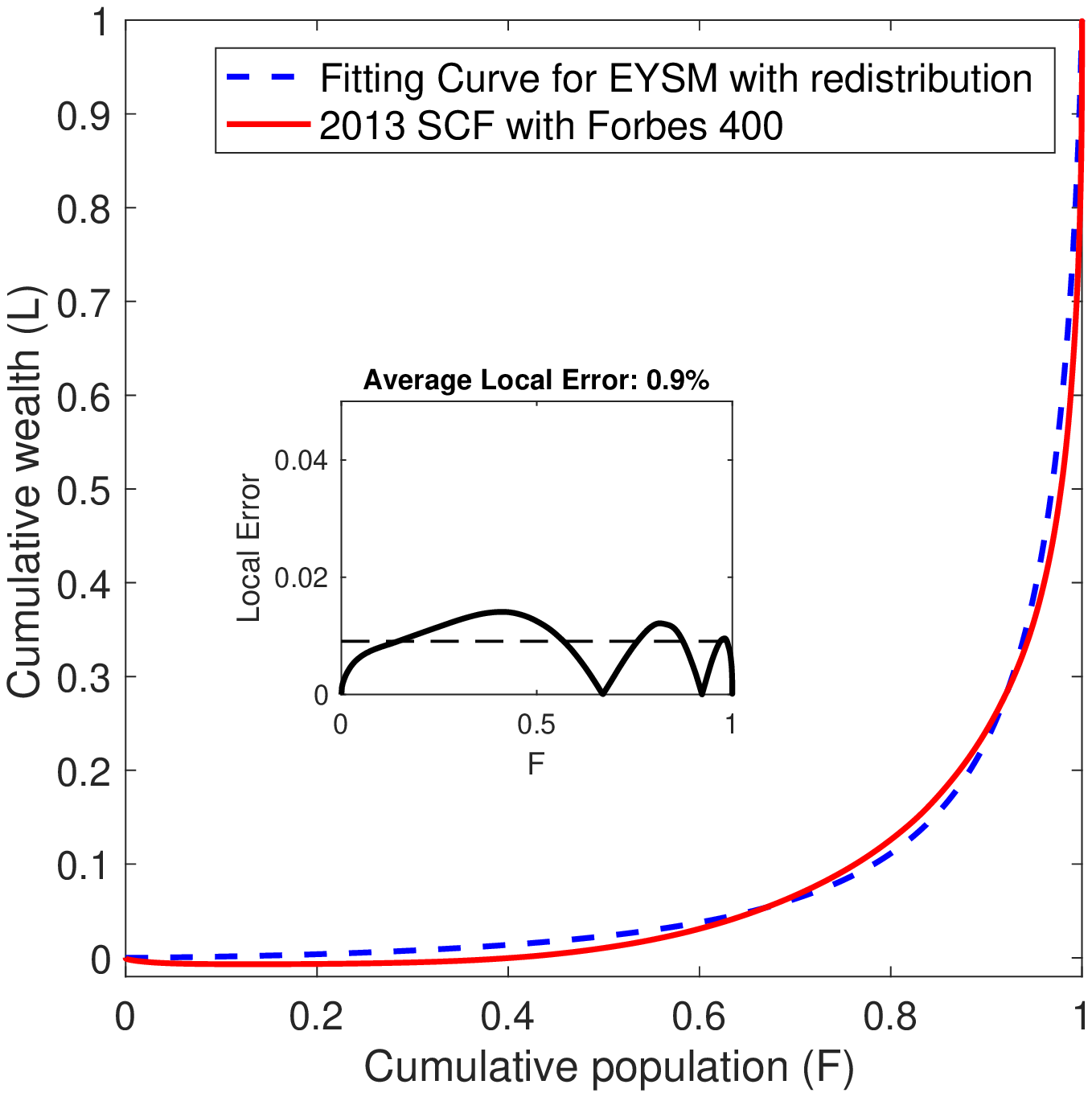}
            \caption[]%
            {{\small EYSM with redistribution}}    
            \label{fig:comp2}
        \end{subfigure}
        \vskip\baselineskip
        \begin{subfigure}[b]{0.475\textwidth}   
            \centering 
            \includegraphics[width=\textwidth]{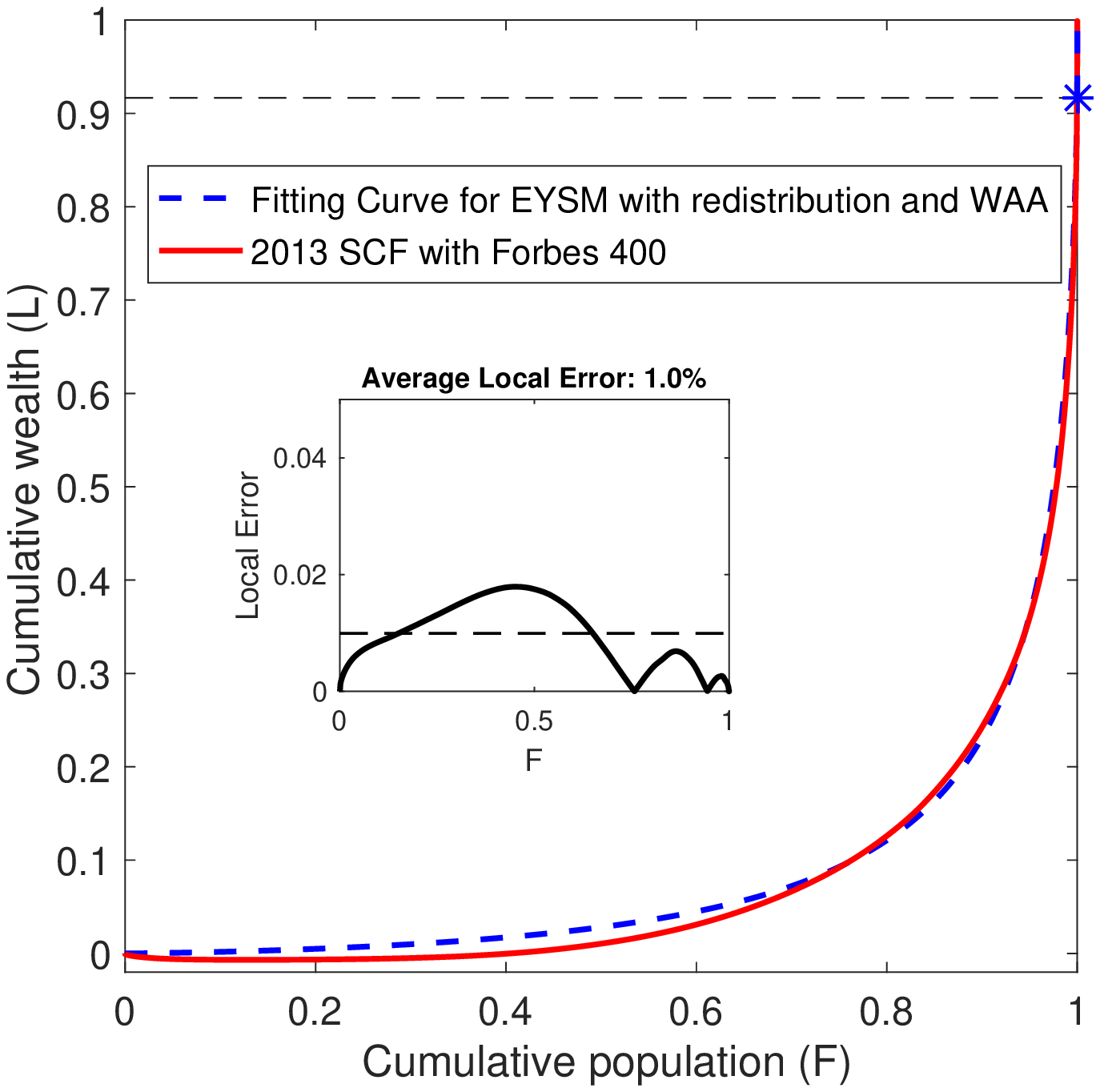}
            \caption[]%
            {{\small EYSM with redistribution and WAA}}    
            \label{fig:comp3}
        \end{subfigure}
        \quad
        \begin{subfigure}[b]{0.475\textwidth}   
            \centering 
            \includegraphics[width=\textwidth]{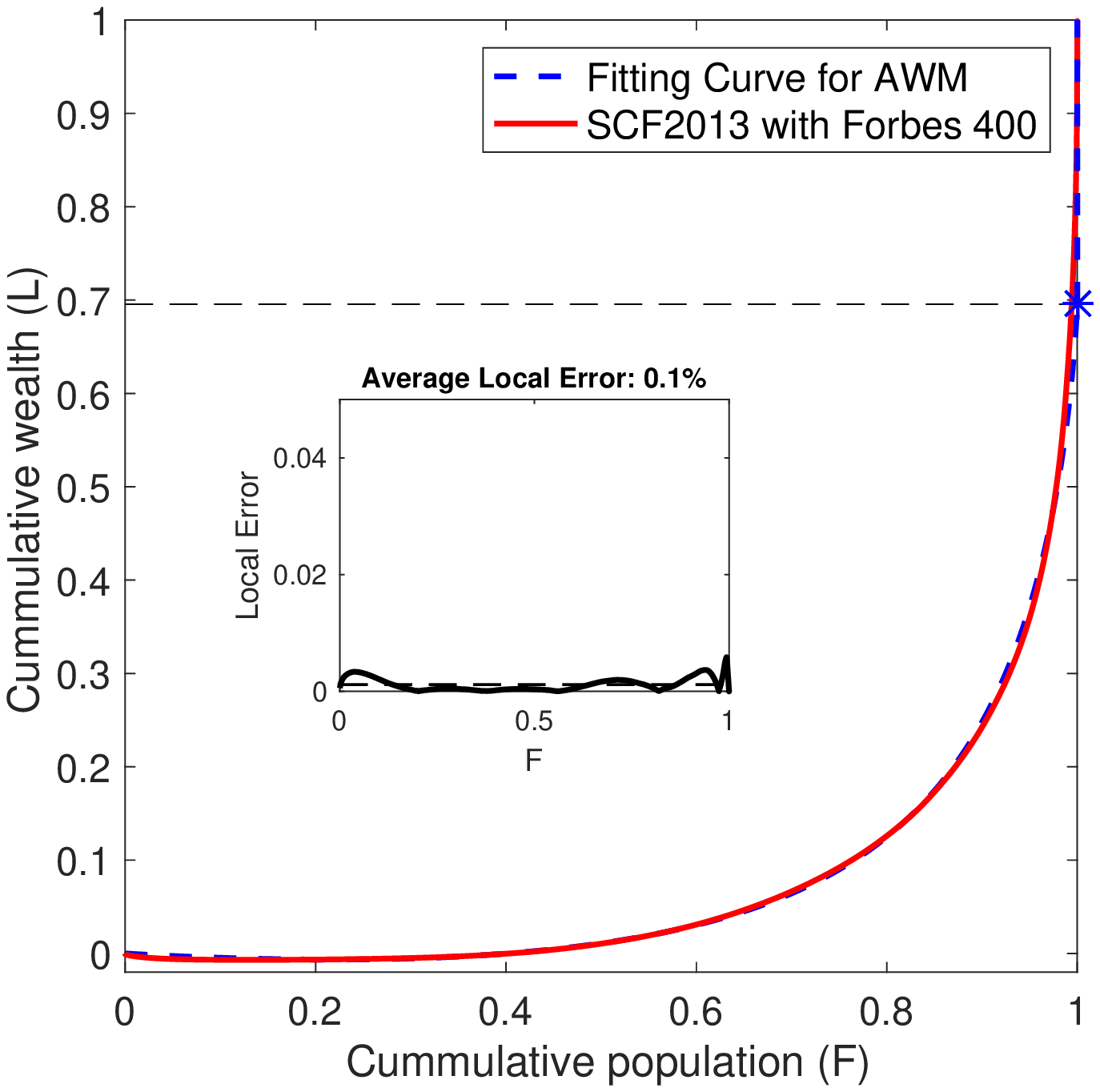}
            \caption[]%
            {{\small Affine Wealth Model}}    
            \label{fig:comp4}
        \end{subfigure}
        \caption{\small Fits of four different models to the empirical Lorenz Curve for the 2013 U.S.\ Survey of Consumer Finances data, with the addition of the Forbes 400.  For each fit, we searched for the optimal parameter vector that minimizes the $L_1$ norm between the empirical and model Lorenz curves.  The local error plotted in the inset was computed as described in Subsection~\ref{ssec:FittingMethod}.  The four fits demonstrate increasing improvement in their ability to fit empirical data, in the order of their presentation.  The fit for the AWM, introduced in this paper, is superior to the other three models in its ability to capture the characteristics of the wealth distribution both in the lower-wealth region (including negative wealth) as well as in the the upper tail.}
        \label{fig:model_compare}
\end{figure*}

In this subsection, we apply the fitting technique of Subsection~\ref{ssec:FittingMethod} to all four models described in Subsection~\ref{ssec:ModelDescription}.  For this purpose, we use the 2013 SCF data, including the Forbes 400, as described in Subsection~\ref{ssec:Data}. All the fitting results are shown in Fig.~(\ref{fig:model_compare}), and all the optimal parameters found as well as the comparisons between the fitting Ginis and the empirical Gini are summarized in Table.(\ref{tab:model_compare})

Fig.~(\ref{fig:comp1}) shows a fitting to the SCF data for the baseline SAM model.  Although the Gini coefficient of the fitting curve is close to the empirical Gini, the discrepancy between the two curves clearly leaves something to be desired. The fit suggests that the SAM is unable to capture the behavior of the actual Lorenz curve both in the lower-wealth region and in the upper tail.  It seriously overestimates the Lorenz curve in the lower-wealth region, far beyond what can be explained by that fact that it does not allow for negative wealth.  Moreover, the SAM's asymptotically power-law tail is seen to badly underestimate the empirical upper tail.  The local error incurred, plotted in the inset, has an average in the vicinity of 2\%.

Fig.~(\ref{fig:comp2}) shows the fitting to the SCF data for the EYSM with redistribution only, as described in Subsection~\ref{ssec:EYSM}.  This is the simplest nonlinear, binary-transaction model that we considered that gives a stable distribution.  Even though there is only a single parameter in this model, namely the redistribution $\chi$, just as there was in the SAM, the extent to which the fit has improved throughout the entire range of $f$ is remarkable. This suggests that nonlinear models with binary transactions have large advantages over linear models. Once again, the largest local error occurs in the low-wealth region, but this time it may well be due to the fact that the model does not allow for negative wealth.

Fig.~(\ref{fig:comp3}) shows the fit to the 2013 SCF data for the EYSM with both redistribution and WAA, as described in Subsection~\ref{ssec:EYSM}.  Recall that this model is capable of wealth condensation.  The result clearly demonstrates that the best fit to the data lies in the supercritical regime, suggesting that the U.S.\ wealth distribution at that time was oligarchical, with 8.33\% of the total wealth of the country held by a vanishingly small number of agents. Note that the upper tail of the fit is significantly improved, as compared to the two previous fits.  Still, the EYSM does not allow for negative wealth, and hence there is still a large discrepancy in the lower-wealth region.

Finally, Fig.~(\ref{fig:comp4}) shows the fitting to the 2013 SCF data for the AWM.  Even a glance at the figure is sufficient to tell that the AWM is better than any of the other three models considered.  Of course, it has more parameters than the others, but three parameters does not seem like a high price to pay for this kind of faithfulness to empirical data.  Because the AWM allows for negative wealth, it is able to capture what is happening in the low-wealth region of the Lorenz curve, yet without losing its accuracy in the upper tail.  The model and empirical curves lie nearly on top of one another, the discrepancy is reduced by nearly an order of magnitude compared with the other fits, and the average local error is reduced to about only one tenth percent. In summary, we feel that the AWM has provided a reasonable way to extend the EYSM to the negative wealth regime, enabling the accurate quantitative modeling of empirical data.

\begin{table*}
\centering
  \begin{tabular}{ | p{4cm} | p{1.5cm} | p{1.5cm} | p{1.5cm} | p{1.5cm} | p{1.5cm} |}
    \hline
    Models & $\chi_{opt}$ & $\zeta_{opt}$ & $\kappa_{opt}$ & Fitting Gini & Empirical Gini\\ \hline \hline
    Single Agent Model & 0.0066 & - & - & 83.29\% & \multirow{4}{*}{85.50\%} \\ \cline{1-5}
    EYSM w/ redist. & 0.016 & - & - & 83.85\% & \\ \cline{1-5}
    EYSM w/ redist. \& WAA & 0.022 & 0.024 & - & 83.76\% & \\ \cline{1-5}
    Affine Wealth Model & 0.046 & 0.064 & 0.076 & 85.59\% & \\ \hline
  \end{tabular}
\caption{Optimal values of the parameters and fitted Ginis found for each model in Fig.~\ref{fig:model_compare}}
\label{tab:model_compare}
\end{table*}

\subsection{Results for SCF from 1989 to 2016}
\label{ssec:Results}

Having established the superiority of the AWM over the other three models considered in Subsection~\ref{ssec:Comparison}, we henceforth restrict our attention to the AWM and examine empirical data from the SCF over the course of time. The SCF has been conducted in three years intervals, from 1989 to 2016, so there are ten plots in total. For each of these ten datasets, we solve the ``inverse problem'' of finding the AWM parameter vector $\theta = \langle\chi,\zeta,\kappa\rangle$ that minimizes the $L_1$ norm of the difference between the empirical and model Lorenz curves.

The results of the fits are shown in Fig.~\ref{fig:time_course}, and all the optimal parameters found as well as comparisons between the fitting Ginis and the empirical Ginis are summarized in Table.(\ref{tab:time_course}). For each year considered, we report the best fittings as well as the optimal parameters.  The plots indicate that the AWM fits to the empirical data remarkably well.  For all the three-parameter fits, the empirical and model Lorenz curves are nearly indistinguishable, and average local errors are in the vicinity of 0.15\%. One key observation is that all fits fall into the supercritical regime with $\chi < \zeta$.  Therefore the Lorenz curves computed from the numerical solutions all hit the right boundary at $((1+\lambda)\chi/\zeta - \lambda, 1)$, as described in Eq.(~\ref{eq:LorenzRightBoundary}), instead of at $(1, 1)$. The horizontal dotted lines indicate where the Lorenz curves hit the boundary, and the model Lorenz curve is a vertical line above this point. This strongly suggests that the U.S.\ wealth distribution has been in a state of partial wealth condensation -- or partial oligarchy -- for all the years of the SCF.  The fraction of the total societal wealth held by the oligarchy is $(1+\lambda)\left(1 - \sfrac{\chi}{\zeta}\right)$, by Eq.~(\ref{eq:OligarchyWealthFraction}). From the plots, we can see that, this fraction is in the vicinity of 20\% to 30\%. 

\begin{figure*}
        \centering
        \begin{subfigure}[b]{0.475\textwidth}
            \centering
            \includegraphics[width=\textwidth]{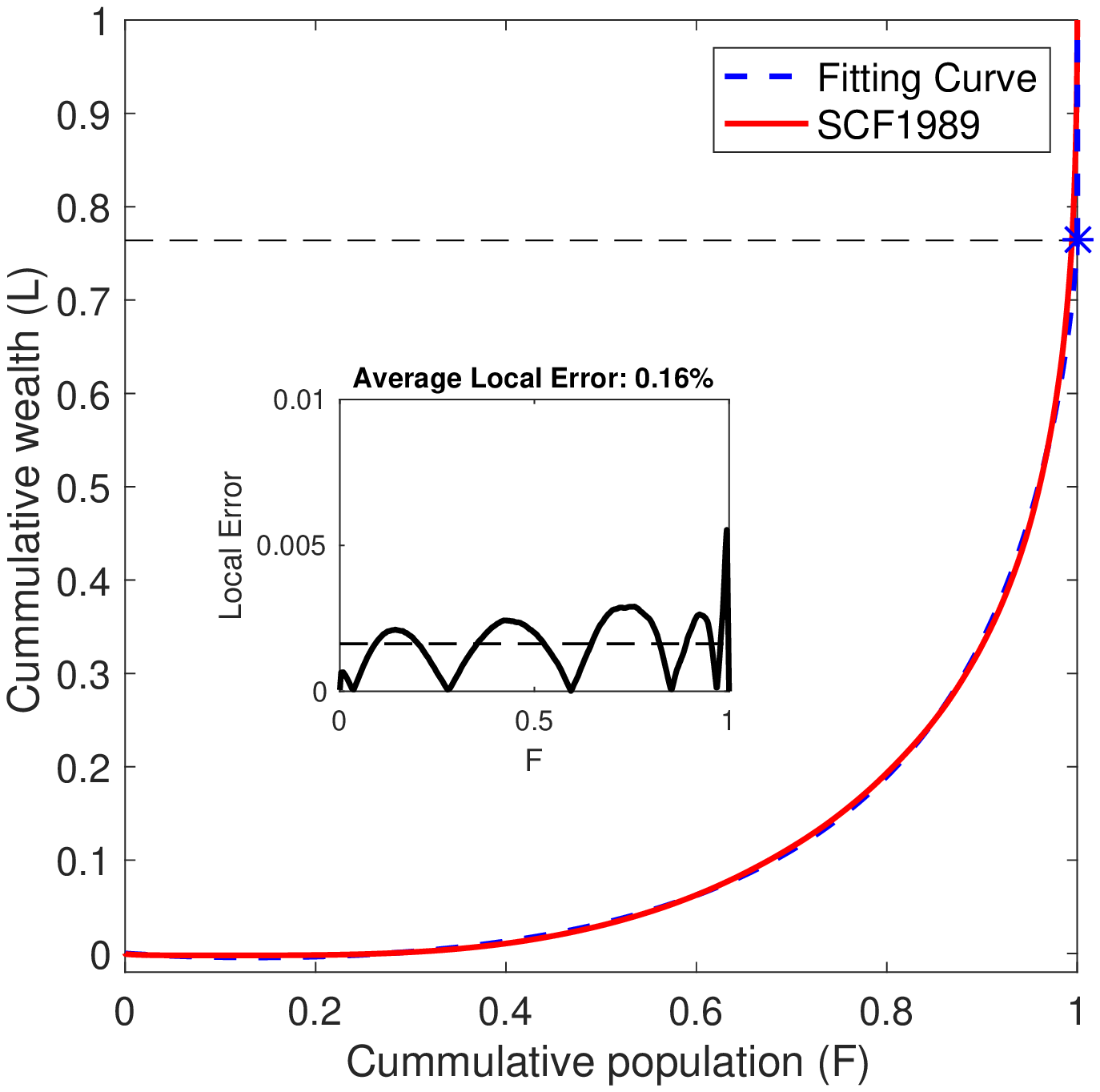}
            \caption{{\small Lorenz Curve Fit of SCF1989}}
            \label{lorenz:1989}
        \end{subfigure}
        \begin{subfigure}[b]{0.475\textwidth}  
            \centering 
            \includegraphics[width=\textwidth]{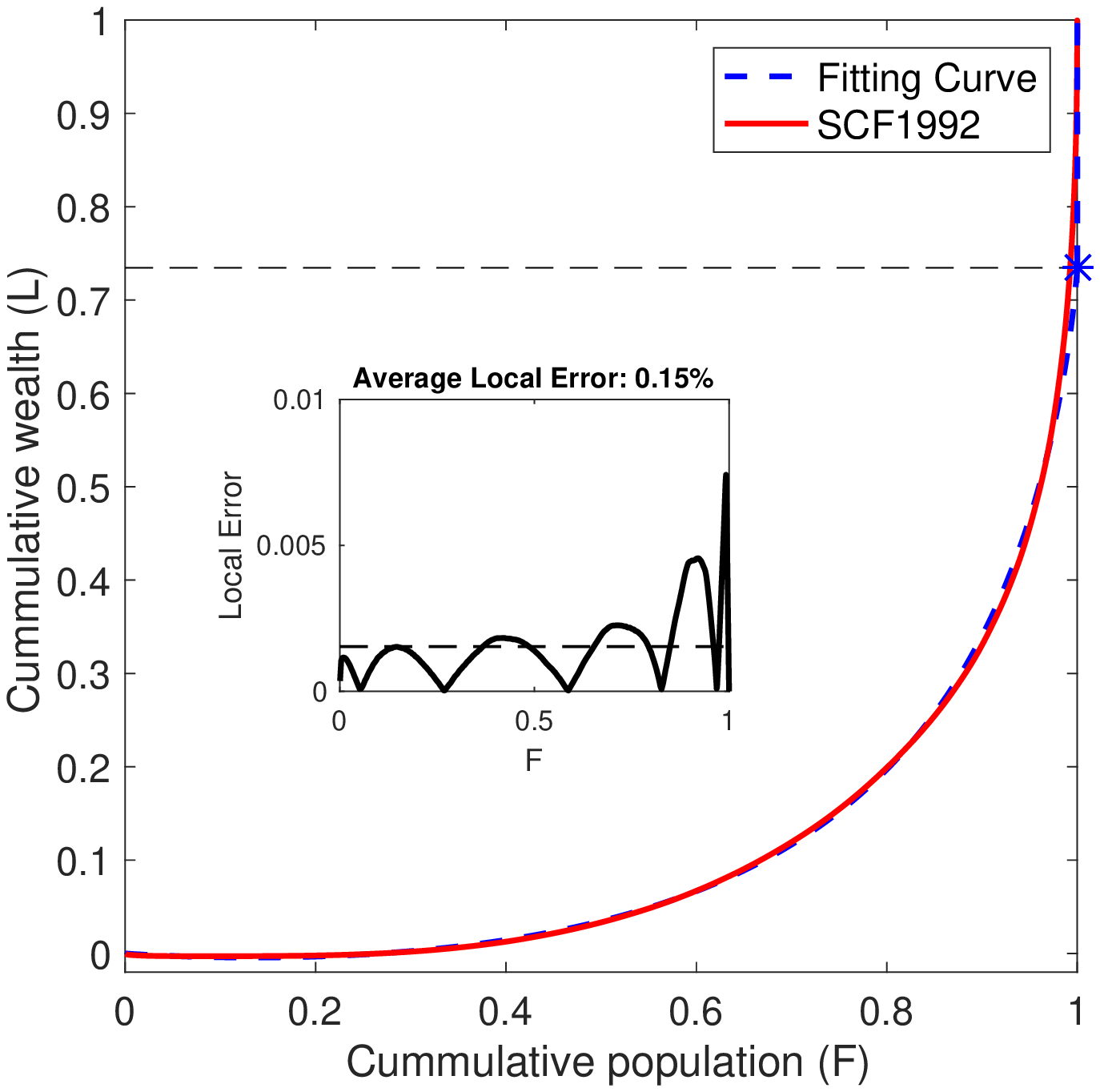}
            \caption{{\small Lorenz Curve Fit of SCF1992}}
            \label{lorenz:1992}
        \end{subfigure}
        
        \begin{subfigure}[b]{0.475\textwidth}   
            \centering 
            \includegraphics[width=\textwidth]{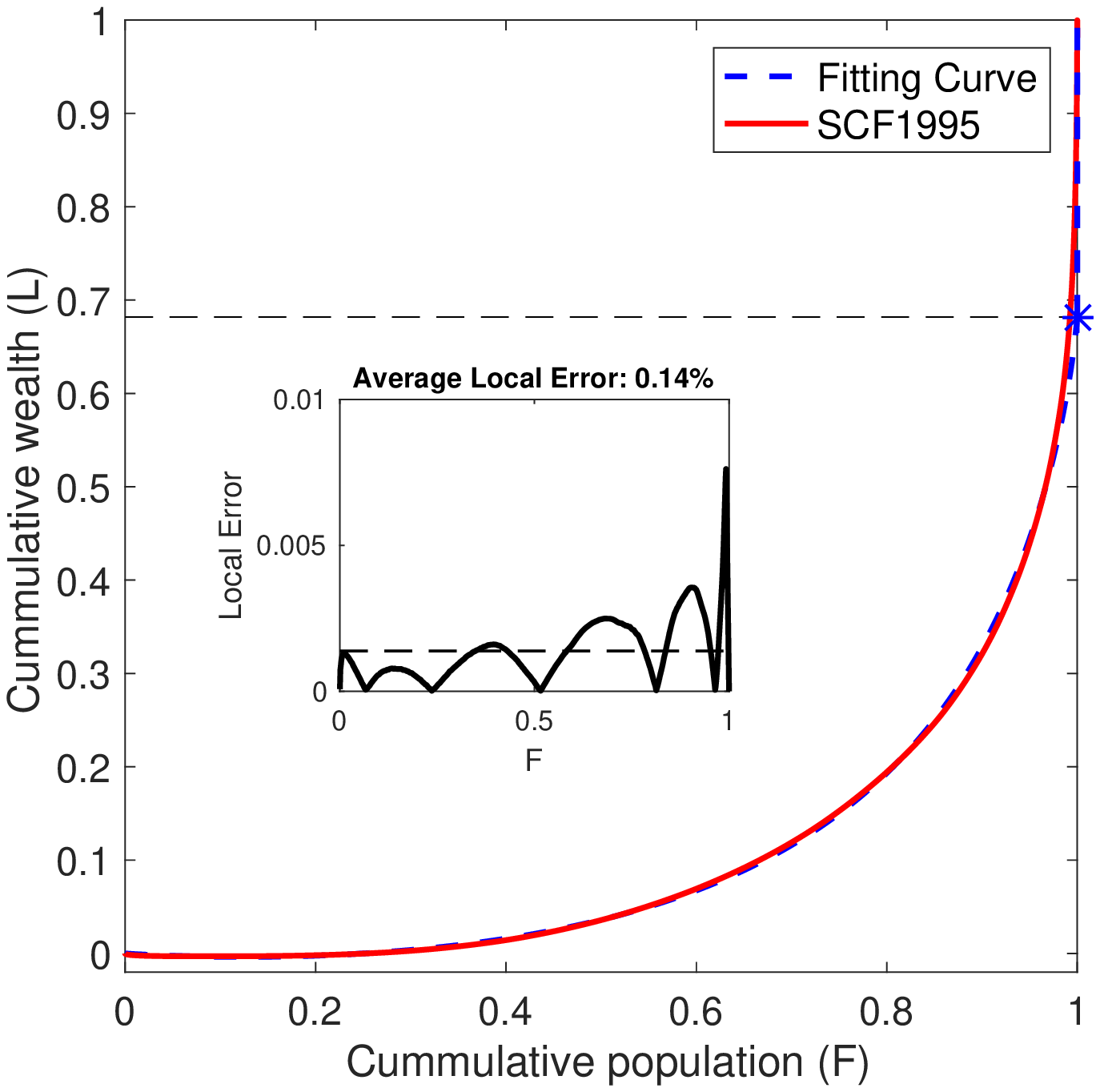}
            \caption{{\small Lorenz Curve Fit of SCF1995}}
            \label{lorenz:1995}
        \end{subfigure}
        \begin{subfigure}[b]{0.475\textwidth}
            \centering
            \includegraphics[width=\textwidth]{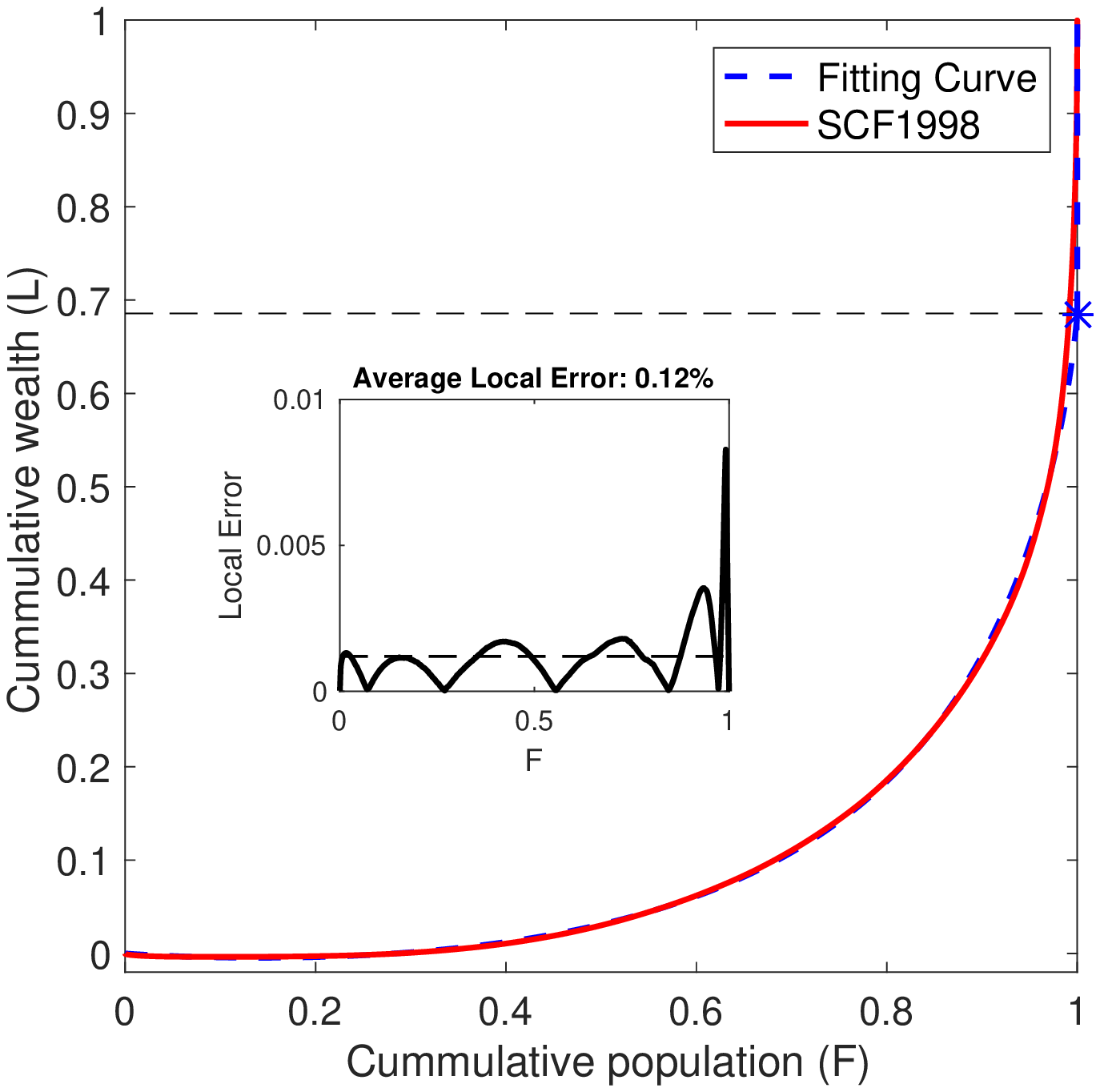}
            \caption{{\small Lorenz Curve Fit of SCF1998}}
            \label{lorenz:1998}
        \end{subfigure}
        
        \begin{subfigure}[b]{0.475\textwidth}  
            \centering
            \includegraphics[width=\textwidth]{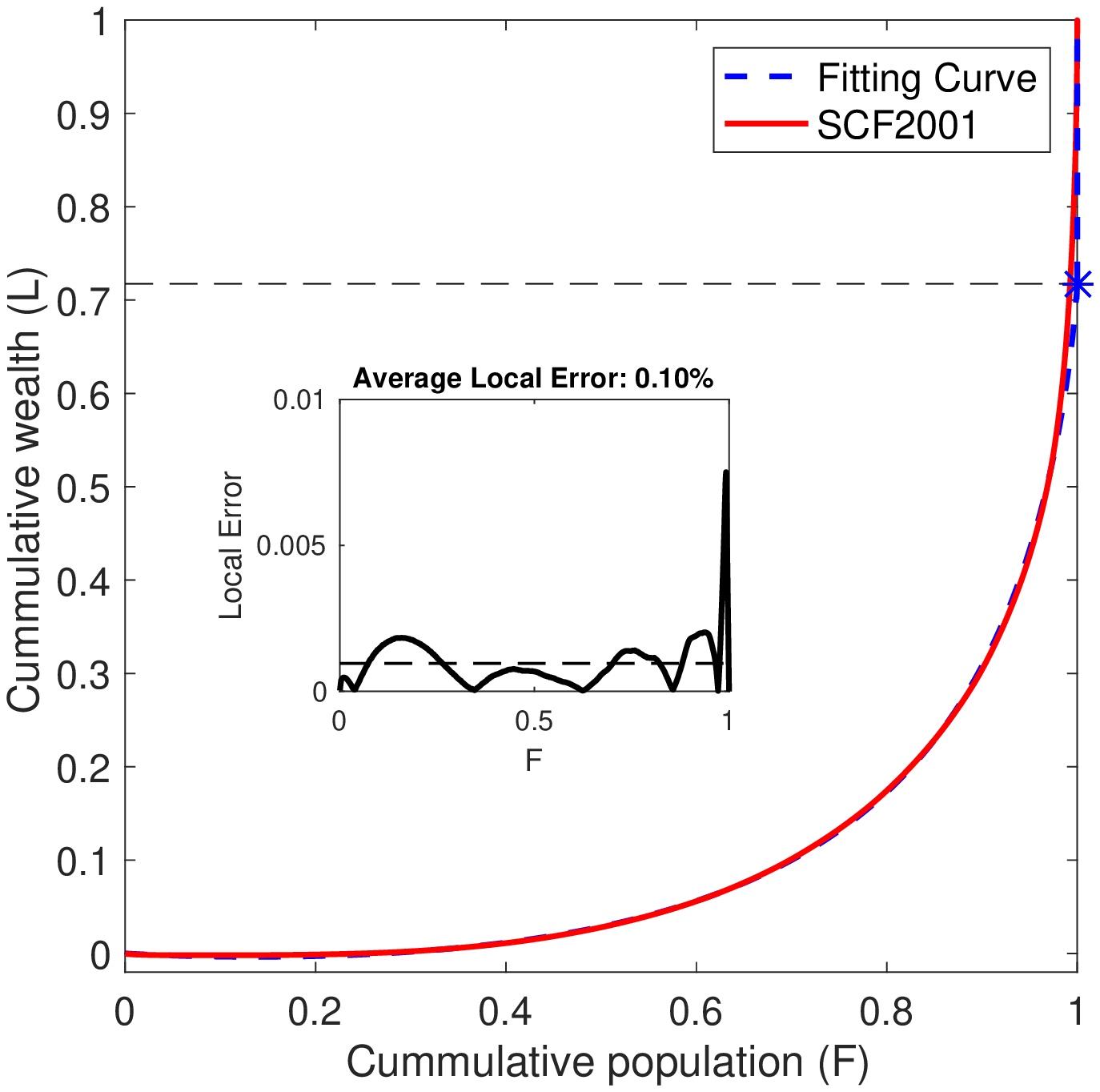}
            \caption{{\small Lorenz Curve Fit of SCF2001}}
            \label{lorenz:2001}
        \end{subfigure}
        \begin{subfigure}[b]{0.475\textwidth}   
            \centering 
            \includegraphics[width=\textwidth]{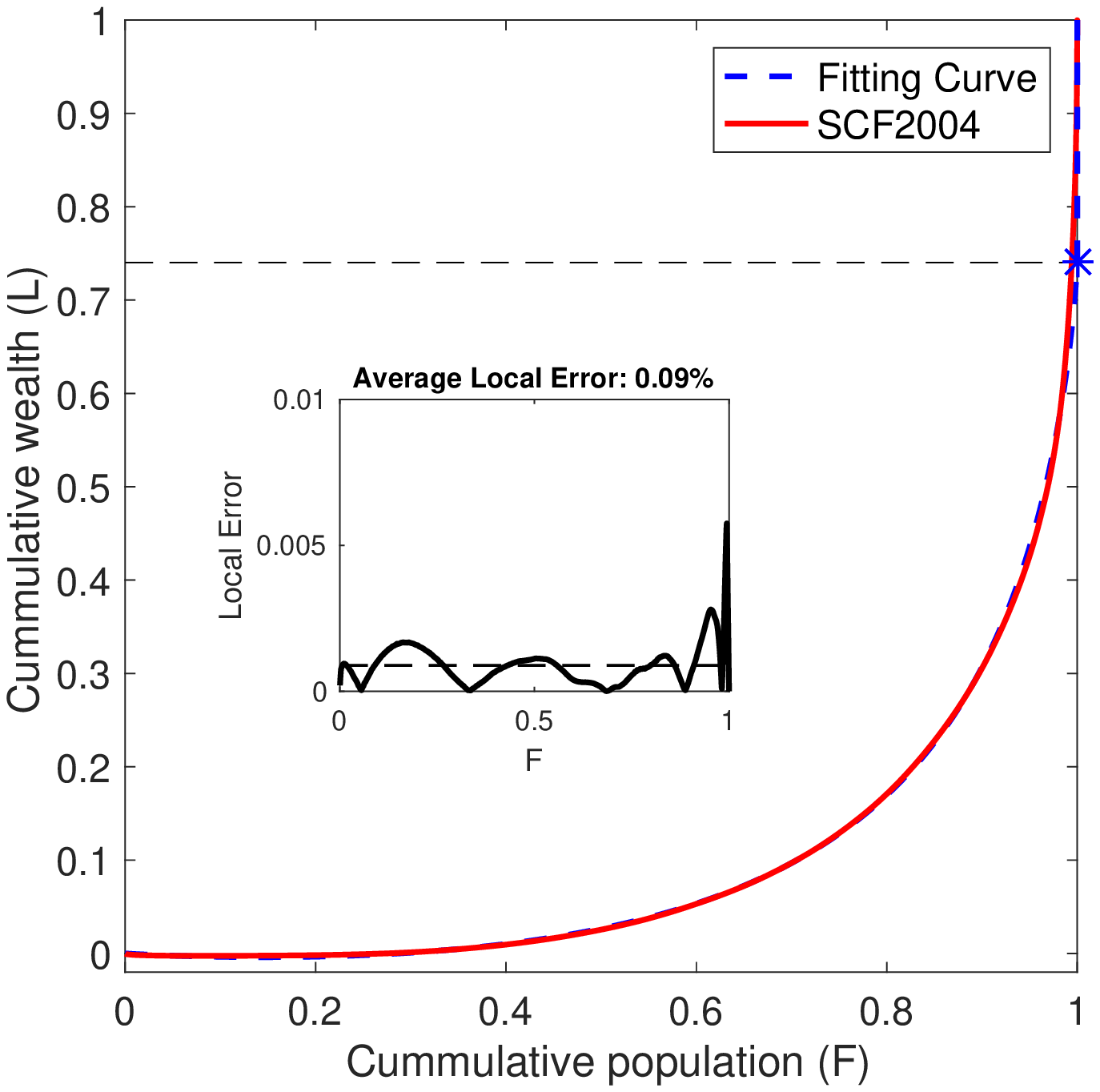}
            \caption{{\small Lorenz Curve Fit of SCF2004}}
            \label{lorenz:2004}
        \end{subfigure}
        \caption{Continues on next page}
%        	\label{fig:time_course}
\end{figure*}

\begin{figure*}
	\centering
	\ContinuedFloat
        \begin{subfigure}[b]{0.475\textwidth}
            \centering
            \includegraphics[width=\textwidth]{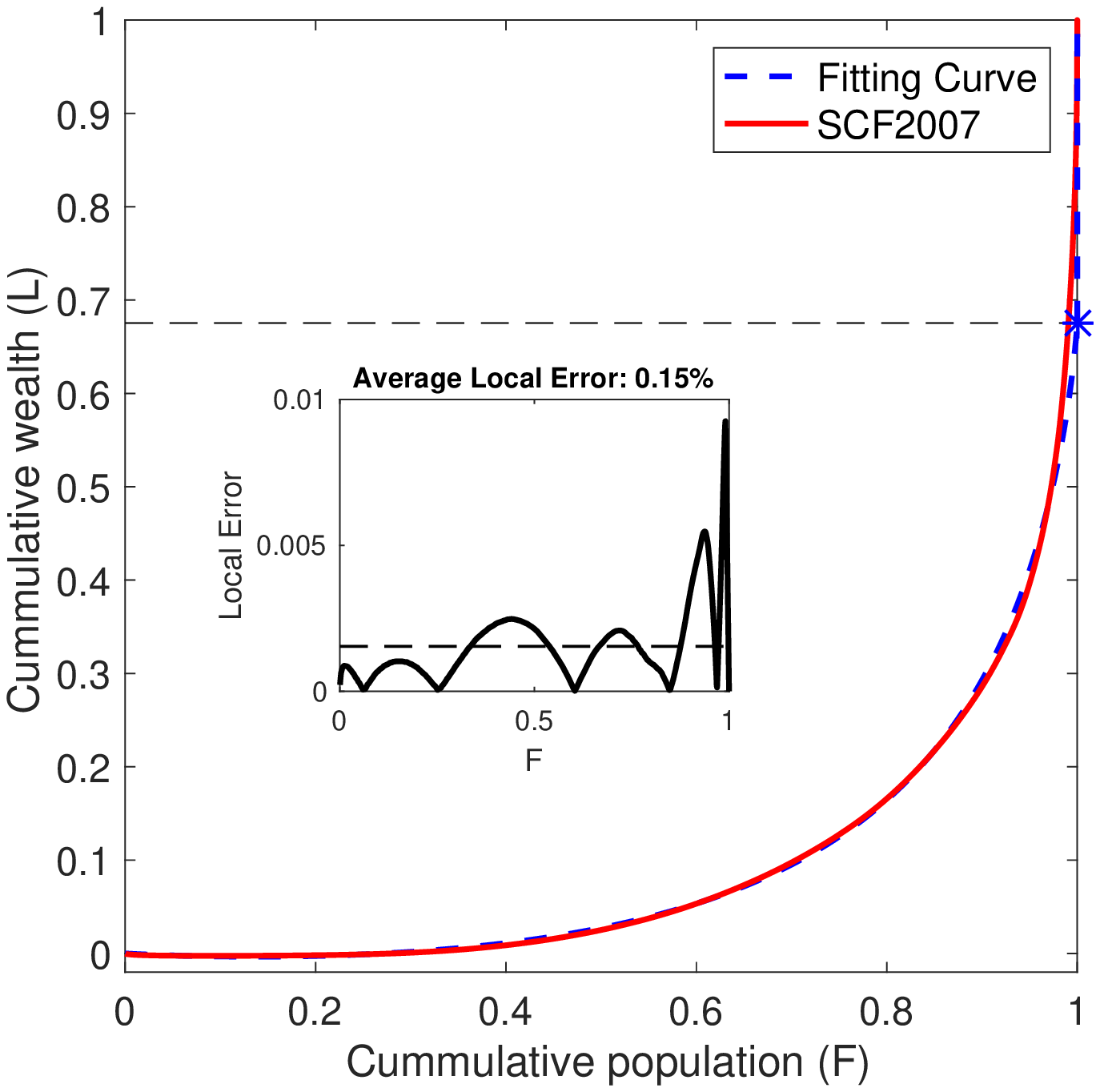}
            \caption{{\small Lorenz Curve Fit of SCF2007}}
            \label{lorenz:2007}
        \end{subfigure}
        \begin{subfigure}[b]{0.475\textwidth}  
            \centering 
            \includegraphics[width=\textwidth]{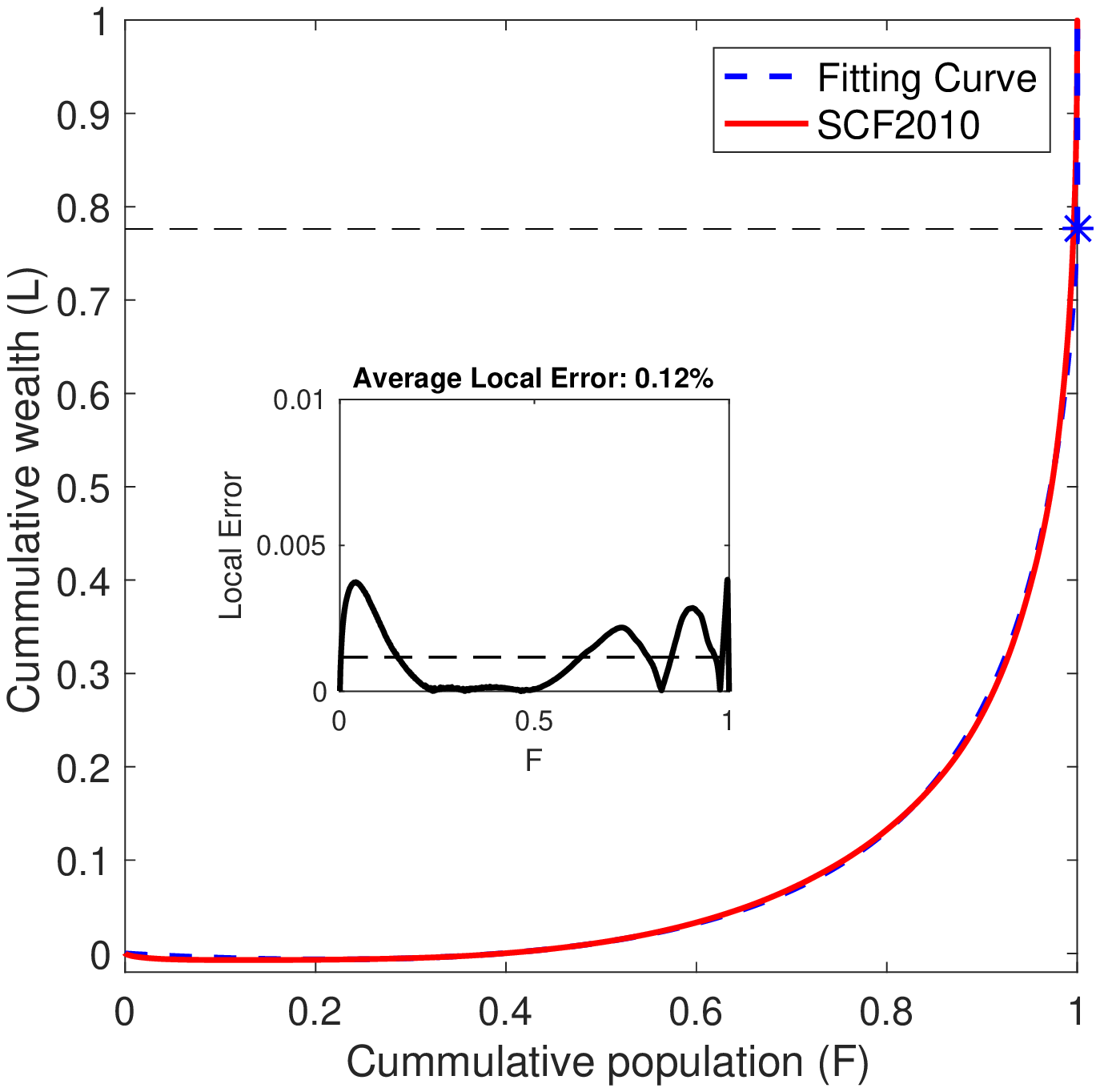}
            \caption{{\small Lorenz Curve Fit of SCF2010}}
            \label{lorenz:2010}
        \end{subfigure}
        
        \begin{subfigure}[b]{0.475\textwidth}   
            \centering 
            \includegraphics[width=\textwidth]{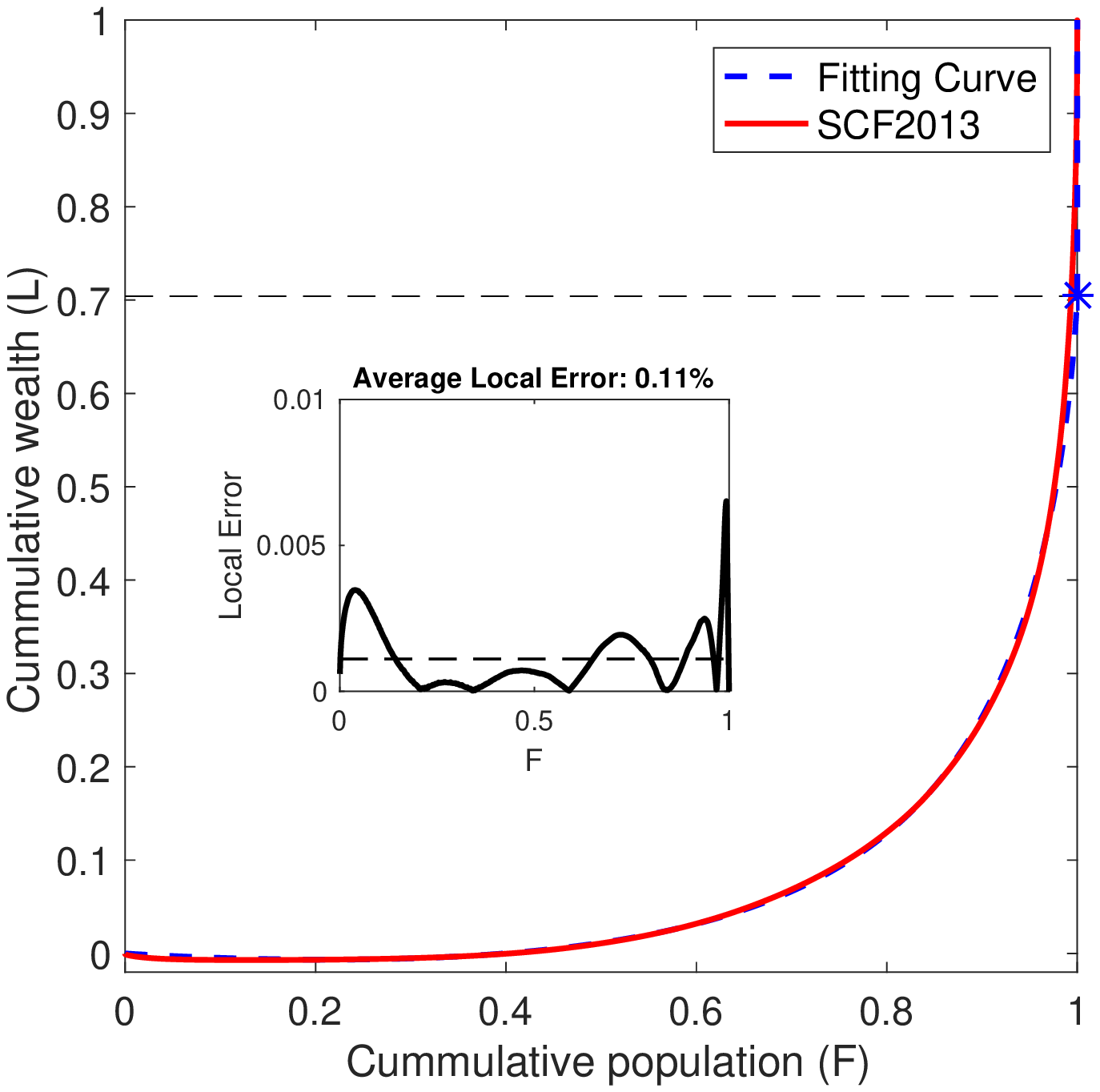}
            \caption{{\small Lorenz Curve Fit of SCF2013}}
            \label{lorenz:2013}
        \end{subfigure}
        \begin{subfigure}[b]{0.475\textwidth}   
            \centering 
            \includegraphics[width=\textwidth]{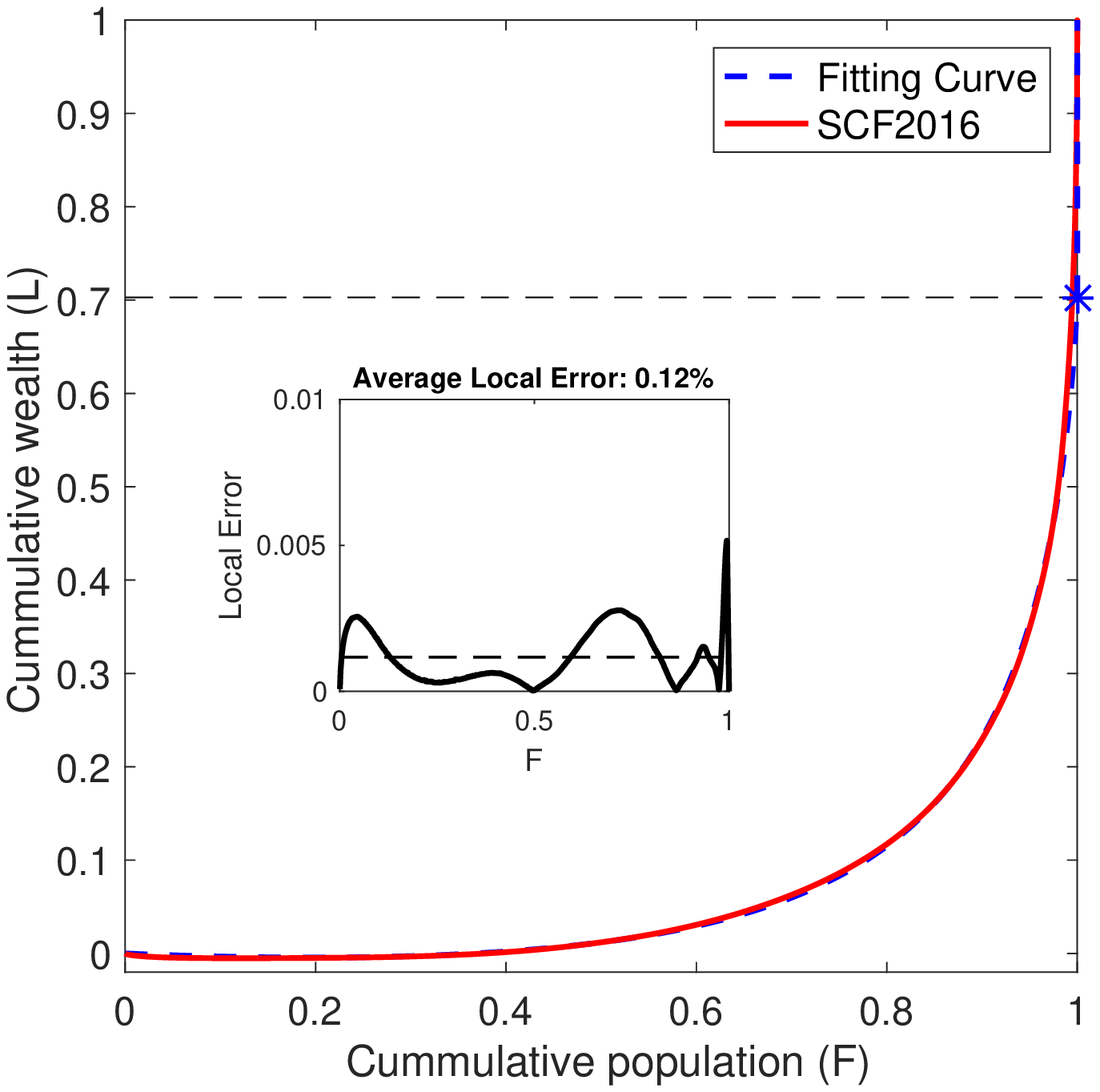}
            \caption{{\small Lorenz Curve Fit of SCF2016}}
            \label{lorenz:2016}
        \end{subfigure}
	\caption{\small Optimal fits of the AWM to SCF data from 1989 to 2016.  For each year, we determined the parameter triplet $\langle\chi,\zeta,\kappa\rangle$ that minimizes the $L_1$ norm of the difference between the empirical and model Lorenz curves.  Our results demonstrate that the AWM fits  the empirical data remarkably well for all ten datasets.  All the fits are in the supercritical regime, which strongly suggests that U.S.\ wealth distribution is partially wealth-condensed; i.e., a finite fraction of the total wealth of the society is held by an infinitesimal fraction of the agents.  This fraction can be estimated by Eq.~(\ref{eq:OligarchyWealthFraction}), which was consistently in the vicinity of 30\% for all the years of the study.} 
	\label{fig:time_course}
\end{figure*}

\begin{table*}
\centering
  \begin{tabular}{ | p{1.5cm} | p{1.5cm} | p{1.5cm} | p{1.5cm} | p{1.5cm} | p{1.5cm} | p{3cm} |}
    \hline
    Years & $\chi_{opt}$ & $\zeta_{opt}$ & $\kappa_{opt}$ & Fitting Gini & Empirical Gini & fraction of wealth held by oligarch\\ \hline \hline
    1989 & 0.088 & 0.112 & 0.092 & 79.17\% & 78.96\% & 23.60\% \\ \hline
    1992 & 0.102 & 0.134 & 0.100 & 78.72\% & 78.59\% & 26.53\% \\ \hline
    1995 & 0.104 & 0.146 & 0.096 & 79.27\% & 79.06\% & 31.82\% \\ \hline
    1998 & 0.096 & 0.134 & 0.098 & 80.20\% & 79.99\% & 31.44\% \\ \hline
    2001 & 0.074 & 0.100 & 0.080 & 80.86\% & 80.54\% & 28.26\% \\ \hline
    2004 & 0.070 & 0.092 & 0.080 & 81.07\% & 80.92\% & 25.99\% \\ \hline
    2007 & 0.070 & 0.100 & 0.076 & 81.81\% & 81.61\% & 32.47\% \\ \hline
    2010 & 0.046 & 0.058 & 0.076 & 84.59\% & 84.56\% & 22.39\% \\ \hline
    2013 & 0.048 & 0.066 & 0.078 & 85.24\% & 85.05\% & 29.58\% \\ \hline
    2016 & 0.036 & 0.050 & 0.058 & 86.18\% & 85.94\% & 29.72\% \\ \hline
  \end{tabular}
\caption{Optimal values of the parameters and fitted Ginis found for each year in Fig.~\ref{fig:time_course}}
\label{tab:time_course}
\end{table*}

The ten Lorenz curves from the different years appear very similar to one another, but the fitting parameters vary from year to year. Since each AWM fitting curve is entirely determined by the three model parameters, we can study the trend of the U.S.\ wealth distribution over time by plotting these optimal values of the parameters versus time.

It should be kept in mind that all of our fits are to {\it steady-state} Lorenz curves of the AWM.  In plotting the AWM fitting parameters versus time, we are therefore supposing that their time variation is {\it adiabatic} in nature; in other words, we suppose that the variation of the fitting parameters is too slow to induce the $\frac{\partial P}{\partial t}$ term in the Fokker-Planck equation to make any significant contribution.

\begin{figure*}
        \centering
        \begin{subfigure}[b]{0.475\textwidth}
            \centering
            \includegraphics[width=\textwidth]{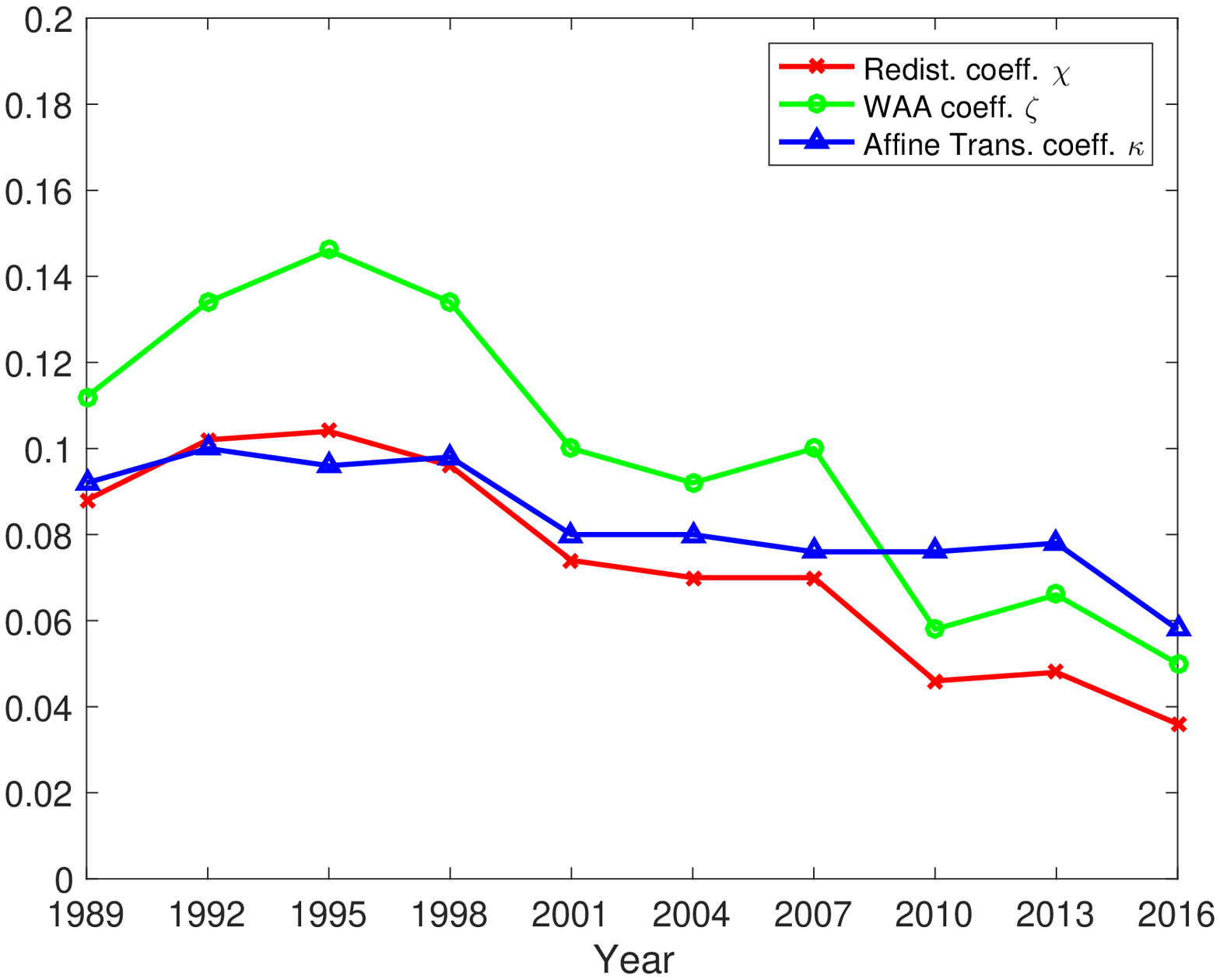}
            \caption{{\small Trend for the three parameters $\langle\chi, \zeta, \kappa\rangle$ in the AWM from 1989 to 2013}}    
            \label{fig:trend1}
        \end{subfigure}
        \begin{subfigure}[b]{0.475\textwidth}  
            \centering 
            \includegraphics[width=\textwidth]{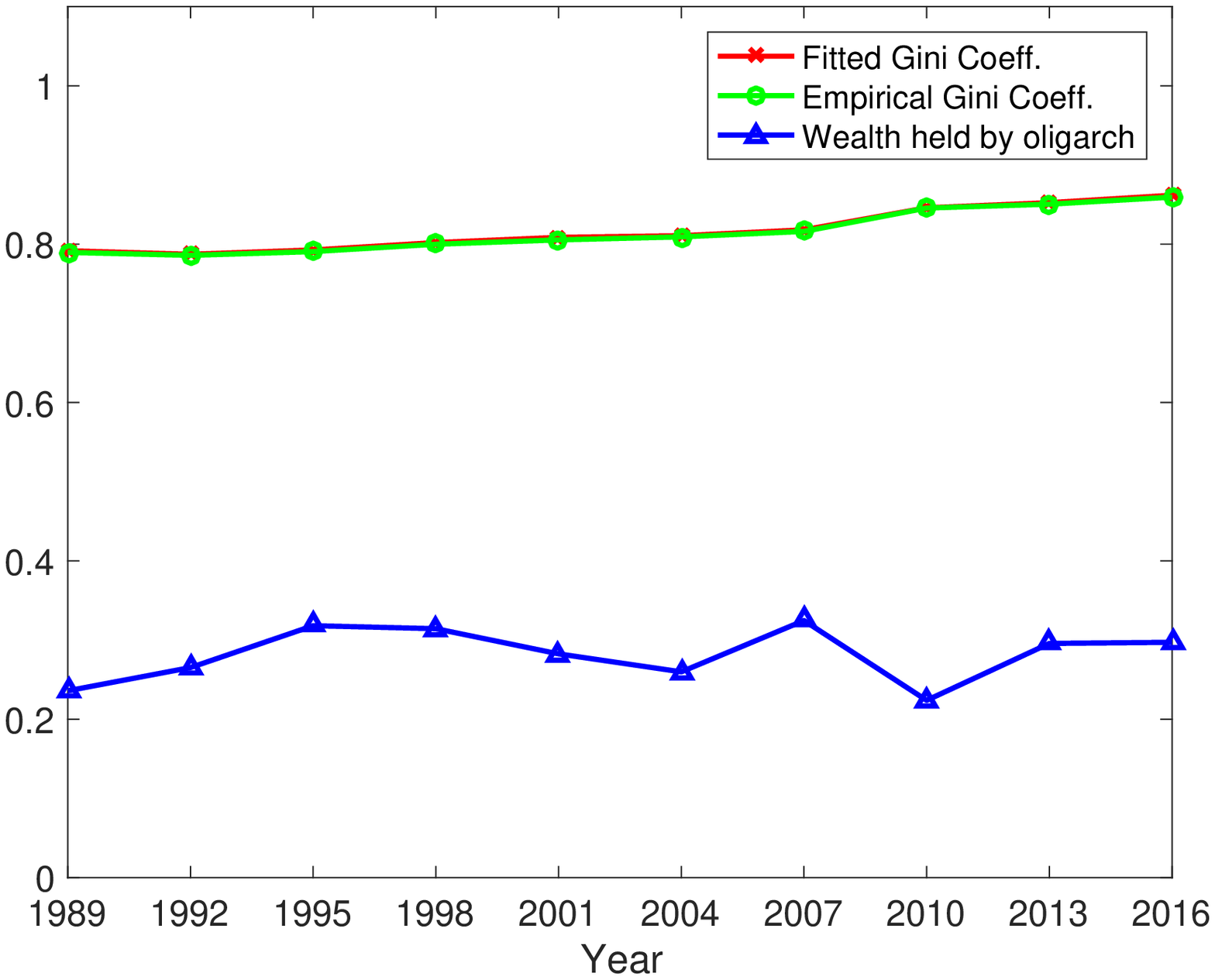}
            \caption{{\small Trend for the Gini coefficient (theoretical and empirical) and the fraction of the total wealth condensed to the oligarchy. from 1989 to 2013}}
            \label{fig:trend2}
        \end{subfigure}
        \caption[]
        {{\small Since the fitting curves are completely determined by the three parameters of the AWM, the change in wealth distribution over time can be summarized by plotting optimal values of these parameters found as a function of time, under the presumption that they change adiabatically.  Our plots demonstrate a correlation between the redistribution parameter $\chi$ and the WAA parameter $\zeta$. The affine transformation coefficient $\kappa$ is less variable than the other two parameters which suggests that the ratio of the lower extreme of the negative-wealth region to the average wealth is relatively stable over the years.  The Gini coefficient of the model data is also very close to that of the empirical data. The fraction of the wealth held by the oligarch can be computed by Eq.~(\ref{eq:OligarchyWealthFraction}), and it is shown that this ratio is relatively stable within the range of 20\% to 30\%.}}
        \label{fig:trend}
\end{figure*}

In Fig.~(\ref{fig:trend1}), we plot the optimal values of the three parameters $\langle\chi,\zeta,\kappa\rangle$ of the AWM, corresponding to the ten Lorenz curves plotted in Fig.~(\ref{fig:time_course}) as a functions of time.  As mentioned earlier, we can confirm that $\chi<\zeta$ throughout this entire period, so the U.S.\ wealth distribution was oligarchical during the entire 27-year period of the study.  It should be pointed out that by ``oligarchical'' here, we mean a very precise thing:  In the space of all valid (classical and distributional) solutions to our Fokker-Planck equation, Eq.~(\ref{eq:fpAWM}), those solutions closest to the SCF data (in the sense that the $L_1$ norm of the discrepancy is smallest) are all distributional solutions exhibiting wealth condensation, i.e., for which $\lim_{f\rightarrow 1^-} \calL(f)<1$.  This gives a mathematically precise definition of the phenomenon of oligarchy.

A second feature we can observe from Fig.~(\ref{fig:trend1}) is that $\kappa$ is much less variable than either $\chi$ or $\zeta$.  Hence the ratio of the lower extreme of the negative-wealth region to the average wealth is relatively constant over the years.

A third feature that is evident from the plot of the parameters versus time is that there seems to be a correlation between $\chi$ and $\zeta$.  This may be because these two parameters are, at least to some extent, redundant in their effect on the agent density function.  Increasing WAA is similar (though certainly not identical) in effect to decreasing redistribution.  Hence, an increase in $\zeta$ can be mitigated to some extent by a simultaneous increase in $\chi$.  It is therefore perhaps not surprising that the ratio of these two parameters, $\chi/\zeta$, is more robust than either one individually.  Since both $\chi/\zeta$ and $\kappa$ are reasonably constant, from Eq.~(\ref{eq:OligarchyWealthFraction}) one would expect the oligarchy wealth fraction to be likewise, and this is verified by the lowest curve in Fig.~(\ref{fig:trend2}), which indicates an oligarchy wealth fraction of between 20\% and 30\% over the course of the study.

The upper curves in Fig (\ref{fig:trend2}) are plots of the empirical and model Gini coefficients. It is unsurprising that these two curves are nearly identical, given the accuracy of our Lorenz curve fits.  Over the 27-year time period of the study, the Gini coefficient has increased from 79\% to 86\%.  This is consistent with figures published by leading economic institutes~\cite{bib:GlobalWealthDataBook, bib:Davies}.

\section{Conclusions and Future Work}
\label{sec:Conclusions}

The AWM, described in this work, provides a new model of wealth distribution that is able to describe empirical data with unprecedented accuracy.  Because the model's parameters are related to specific features of its agent-level description, the trends of these parameters in time, as shown for example in Fig.~(\ref{fig:trend}), enable us to glimpse underlying mechanisms for wealth distribution evolution.  The AWM thereby, at least to some extent, bridges the gap between microeconomics and macroeconomics.  The economic turbulence of 2008, for example, is clearly reflected in a sharp downward movement of the agent-level parameters $\chi$ and $\zeta$, accompanied by a pronounced upward movement in the Gini coefficient -- the latter arguably being a macroeconomic indicator.  For all of the above reasons, we feel that the approach shows great promise, though we acknowledge that a precise relationship between the AWM model parameters and more conventional economic indicators will probably require the involvement of economists, political scientists and public policy specialists to sort out.

Another observation that warrants future study is the nature of the tail of the agent density function, $P(w)$.  It has been demonstrated that the non-oligarchical part of the EYSM agent density function for positive $\chi$ and nonnegative $\zeta$ has a gaussian tail~\cite{bib:Boghosian2017, bib:RBGThesis}.  This seems at least somewhat at odds with -- if not in outright contradiction to -- the conventionally held belief, dating back to Pareto~\cite{bib:Pareto}, that wealth distributions have power-law tails.  We feel that there are a number of good reasons to question this conventional belief.  First, while there seems to be solid empirical evidence for a power-law tail for income distributions~\cite{bib:Yakovenko}, much less work has been done for wealth distributions, perhaps owing to the relative paucity of available data~\footnote{As of this writing, only about twenty countries in the world directly collect wealth data on their household surveys.}.  Second, while the tail of EYSM distributions is always gaussian, the midrange in the limit of small $\chi$ has been shown (albeit only for the case $\zeta=0$) to be nearly power-law in nature~\cite{bib:RBGThesis}. Since our fitted values of $\chi$ are in the vicinity of 4\% to 10\%, this suggests that it is probably very easy to confuse EYSM distributions with power laws, especially if most of one's data is in the midrange -- as is obviously always the case.  Third, we have shown that it is mathematically important to separate the oligarchy, which is best described by distribution theory or by nonstandard analysis in the continuum limit~\cite{bib:DevittLeeWangLiBoghosian}, from the tail of the classical part of the agent density function, since conflating these will make the the tail seem longer than it really is.  This requires delicate numerical analysis which no prior work would have had the motivation to adopt.  Because the accuracy of our fits in Fig.~\ref{fig:time_course} speak for themselves, we believe that it is time to take a fresh look at the actual empirical evidence underlying the long-held belief in power-law tails of wealth distributions.

There is ample room for future work in this area.  As mentioned earlier, our fits of the empirical data were made to steady-state solutions of the Fokker-Planck equation for the AWM, Eq.~(\ref{eq:fpssAWM}).  Yet the optimal model parameters $\langle\chi,\zeta,\kappa\rangle$ found and plotted in Fig.~\ref{fig:trend} are changing in time.  This is no doubt because levels of redistribution, WAA and extreme poverty, respectively, change in time due to public policy and political decisions.  Our approach is nonetheless valid under the assumption that the changes in these parameters are slow enough to be adiabatic in nature -- i.e., not rapid enough for the time derivative, $\frac{\partial P}{\partial t}$, to become appreciable.  Still, future work might focus on taking the time evolution into account.  For this purpose, a more sophisticated fit would be necessary, including time as an independent variable, and fitting to a parameter vector that is a function of time, probably with some smoothness conditions.  This would be a much more difficult fit, and we leave it to future work.

\section{Acknowledgements}
\label{sec:Ack}

We would like to thank the other members of the Wealth Inequality Research Group in the Department of Mathematics at Tufts University for helpful conversations. One of us (BMB) would like to acknowledge the support of the Fondation Math\'{e}matique Jacques Hadamard, as well as the hospitality of the Department of Mathematics at the Universit\'{e} Paris-Sud in Orsay, France, where a first draft of this paper was written.  Additional helpful conversations are also acknowledged at the Department of Economics at the University of Massachusetts at Amherst, and at the Economic Research Group of the Central Bank of Armenia.

\end{document}